\documentclass[11pt,onecolumn,nofootinbib,aps,superscriptaddress,prd]{revtex4}
\usepackage{graphicx,dcolumn,bm,amsfonts,amsmath,color,xcolor,epsfig}
\usepackage{slashed}
\usepackage{graphics}
\usepackage{tikz}
\usepackage{multirow}
\usepackage{amsmath}
\usepackage{gensymb}
\usepackage{epstopdf}
\usepackage{hyperref}
\usepackage{amssymb}
\allowdisplaybreaks

\begin{document}

\title{Investigating the mass spectra of $1F$-wave singly heavy $\Sigma_{Q}$, $\Xi^{\prime}_{Q}$, and $\Omega_{Q}$ baryons}
\author{Ji-Si Pan}
\email{panjisi@gxstnu.edu.cn}
\address{School of Physics and Information Engineering, Guangxi Science $\&$ Technology Normal University,
Laibin 546199, China}
\author{Ji-Hai Pan\footnote{Corresponding author}}
\email{panjihai@lzhit.edu.cn}
\address{Department of Mathematics and Physics, Liuzhou Institute of Technology, Liuzhou 545616, China}

\begin{abstract}
In this work, we enumerated the mass spectra of the experimentally unobserved $1F$-wave states with the the orbital angular momentum $L$ = 3 for the singly heavy $\Sigma_{Q}$, $\Xi^{\prime}_{Q}$, and $\Omega_{Q}$ $(Q=c, b)$ baryons in the framework of quark-diquark configuration using the Regge trajectory model and the scaling rules. To determine the effective masses of the heavy quark and two light quarks, the relativistic effective mass formula are employed by combining the Coulomb potential. Within the spin-dependent Hamiltonian, we construct the mass shift forms as a non-diagonal symmetric $6\times 6$ matrix for the $1F$-wave states of $\Sigma_{Q}$, $\Xi^{\prime}_{Q}$, and $\Omega_{Q}$ baryons. Our analysis of mass spectra provides valuable insights to guide future experimental investigations, and enhances the understanding of the spectroscopic properties of  unobserved $1F$-wave orbital excitations for these singly heavy baryons.
\end{abstract}

\maketitle

\section{introduction}\label{Sec.I}

Singly heavy baryons are composed of the heavy quark $(Q)$ with two light quarks $(qq)$, where $Q$ = $c$ or $b$, and $q$ = $u$, $d$ or $s$. The two light quarks can couple to a total spin of $S_{qq}$ = 1. In recent years, several new singly heavy baryons have been experimentally observed and extensively studied, yielding a wealth of precise data. The study of the mass spectra of singly heavy baryons can help us understand the strong interactions within the internal structure of hadrons, thereby deepening our understanding of quantum chromodynamics (QCD) in the non-perturbative regime.

For the $1S$-wave states of singly heavy baryons, except the $\Omega_{b}$ baryon, the remaining have been observed experimentally. The measured masses are listed in Table \ref{Table 1} with spin parity quantum numbers $J^{P}$. In the charm baryons, $\Sigma_{c}(2455)$ and $\Sigma_{c}(2520)$, which can be well-interpreted as $S$-wave states from Particle Data Group (PDG) \cite{Navas:PP888} with $J^{P}=1/2^{+}$, and $J^{P}=3/2^{+}$, respectively. In Refs. \cite{Averyet:PPP888} and \cite{Gibbonset:PPP888}, CLEO Collaboration reported two states $\Xi^{0}_{c}$ and $\Xi_{c}(2645)$ in the decay channels $\Xi^{+}_{c}\pi^{-}$ and $\Xi^{0}_{c}\pi^{+}$ as $S$-wave states with $J^{P}=1/2^{+}$, $J^{P}=3/2^{+}$, respectively. There are two ground states for $\Omega_{c}$ baryons, defined as $\Omega_{c}^{0}$ and $\Omega_{c}(2770)^{0}$, with $J^{P}=1/2^{+}$ and $J^{P}=3/2^{+}$, respectively, have been discovered in Refs. \cite{Navas:PP888, AubertB:A11}. Note that only one ground state for $\Omega_{b}$ baryons at present, D0 Collaboration report the first observation of the $\Omega_{b}^{-}$ baryon in Ref. \cite{AbazovV:A11}, fully reconstructed from its decay $\Omega_{b}^{-}$ $\rightarrow$ $J/\psi\Omega^{-}$. In the bottom baryons, for the ground states of the $\Sigma_b$ baryons, $\Sigma_b(5815)$ and $\Sigma^*_b(5835)$ states have been observed by the CDF Collaboration in Ref. \cite{collaborationT:PPP888}, long after the previous discovery. Then, two new charged  $\Xi^{\prime}_{b}(5935)$ and $\Xi^{\ast}_{b}(5955)^{-}$ states were reported by LHCb in 2015 \cite{Aaijolla:PPP888} from Run 1 data. A year later, $\Xi_{b}(5945)$ with $J^{P}=3/2^{+}$, which was confirmed by LHCb with a refined mass measurement \cite{RAaijolla:PPP888}.

Additionally, LHCb Collaboration recently discovered five new narrow $\Omega_{c}$ states observed in decay channel $\Xi^{+}_{c}K$ \cite{Aaij:A11}: $\Omega_{c}(3000)^{0}$, $\Omega_{c}(3050)^{0}$, $\Omega_{c}(3065)^{0}$, $\Omega_{c}(3090)^{0}$, and $\Omega_{c}(3120)^{0}$. Later, the Belle Collaboration confirmed the existence of these excited $\Omega_{c}$ states \cite{YeltonBel:PP888}. These five states can be good $P$-wave charmed baryon candidates. In 2020, LHCb Collaboration reported the discovery of four narrow excited $\Omega_{b}$ states in the decay channel $\Xi^{0}_{b}K^{-}$ \cite{Aaij:A12}: $\Omega_{b}(6316)^{-}$, $\Omega_{b}(6330)^{-}$, $\Omega_{b}(6340)^{-}$, $\Omega _{b}(6350)^{-}$. More recently in 2023, $\Omega_{c}(3185)^{0}$ and $\Omega_{c}(3327)^{0}$ states of $\Omega_{c}$ baryons were also observed by LHCb Collaboration \cite{Aaij:A13}, but their spin parity quantum numbers have not yet been determined. So far, just one excited state of $\Sigma_{c}$ baryon, $\Sigma_{c}(2800)$, has been discovered by Belle collaborations in the channel of $\Lambda_{c}^{+} \pi$ \cite{Mizuke:PPP888}. Soon its mass was accurately measured by the BaBar Collaboration \cite{BAubertBa:PPP888}. In 2020, LHCb observed three excited $\Xi^{0}_{c}$ resonances called $\Xi_{c}(2923)$, $\Xi_{c}(2939)$, and $\Xi_{c}(2965)$ in the $\Xi^{+}_{c}K^{-}$ mass spectrum \cite{RAaij:A14}, whose masses are very close to each other, see Table \ref{Table 1}. At present, we can not determine their spin parity quantum numbers $J^{P}$ accurately. However, these states may be also good candidates for some of these missing $P$-wave baryon states. Therefore, based on these experimental data, we performed a analysis of the mass spectra for the $1F$-wave singly heavy baryons to provide theoretical support for the discovery of these particles.

Although several excited singly heavy baryons have been observed, their internal structures and quantum number assignments are not yet fully established, especially for higher orbital excitations. In this work, we therefore focus on the mass spectra of the $1F$-wave states for the $\Sigma_{Q}$, $\Xi^{\prime}_{Q}$, and $\Omega_{Q}$ baryons. According to the PDG \cite{Navas:PP888}, the $1S$- and $1P$-wave states of these baryons have already been observed experimentally, while the $1D$- and $2S$-wave states have candidates or theoretical discussions. By contrast, the $F$-wave states of the $\Sigma_{Q}$, $\Xi^{\prime}_{Q}$, and $\Omega_{Q}$ baryons have received considerably less theoretical attention, and the $1F$-wave states baryons remain experimentally completely unknown. This is especially interesting because it contains six states with different total momentum quantum number, several of which can mix through spin-dependent interactions. As a result, a systematic study of these states is still lacking, and provides a useful test of the mass formula and the spin-coupling scheme used in the present model.

\renewcommand\tabcolsep{0.6cm}
\renewcommand{\arraystretch}{0.8}
\begin{table}[htbp]
\caption{Masses and $J^P$ of $\Sigma_{Q}$, $\Xi^{\prime}_{Q}$ and $\Omega_{Q}$ from PDG with our predicted.   \label{Table 1}}
\begin{tabular}{ccccc}
\hline\hline
State                   &Mass (MeV)                &     $J^{P}$   & Ours   &     $J^{P}$ Proposed   \\
\hline
$\Sigma _{c}(2455)^{+}$  &2452.65                  &       $\frac{1}{2}^{+}$     &2452.37$\pm$ 5.07     &       $\frac{1}{2}^{+}$                  \\
$\Sigma _{c}(2520)^{+}$  &2517.4                   &       $\frac{3}{2}^{+}$     &2517.01$\pm$ 2.91     &       $\frac{3}{2}^{+}$                  \\
$\Sigma _{c}(2800)^{+}$  &2792                     &       $?^{?}$               &2791.12$\pm$ 1.87     &       $\frac{3}{2}^{-}$         \\
$\Sigma _{b}^{+}      $  &5810.56$\pm$ 0.25        &       $\frac{1}{2}^{+}$     &5814.42$\pm$ 2.32     &       $\frac{1}{2}^{+}$                   \\
${\Sigma _{b}^{\ast}}^{+} $ &5830.32$\pm$ 0.27     &       $\frac{3}{2}^{+}$     &5830.51$\pm$ 2.08     &       $\frac{3}{2}^{+}$                   \\
$\Sigma _{b}(6097)^{-} $ &6098.0$\pm$ 1.7$\pm$0.5  &       $?^{?}$               &6098.57$\pm$ 2.29     &       $\frac{3}{2}^{-}$         \\
${\Xi _{c}^{\prime}}^{0} $ &2578.7$\pm$ 0.5        &       $\frac{1}{2}^{+}$     &2579.77$\pm$ 4.07     &       $\frac{1}{2}^{+}$                   \\
$\Xi _{c}(2645)^{0}   $ &2646.16$\pm$ 0.25         &       $\frac{3}{2}^{+}$     &2629.28$\pm$ 2.43     &       $\frac{3}{2}^{+}$                   \\
$\Xi _{c}(2923)^{0}   $ &2923.2$\pm$ 0.4           &       $?^{?}$               &2909.34$\pm$ 1.78     &       $\frac{3}{2}^{-}$         \\
$\Xi _{c}(2930)^{0}   $ &2938.55$\pm$ 0.30         &       $?^{?}$               &2937.35$\pm$ 1.81     &       $\frac{3}{2}^{-}$           \\
$\Xi _{c}(3123)^{+}   $ &3122.9$\pm$ 1.3$\pm$0.3   &       $?^{?}$               &3138.13$\pm$ 2.00     &       $\frac{1}{2}^{+}$           \\
$\Xi^{\prime} _{b}(5935)^{-}   $ &5934.9$\pm$ 0.4  &       $\frac{1}{2}^{+}$     &5937.83$\pm$ 8.04     &       $\frac{1}{2}^{+}$                    \\
$\Xi _{b}(5955)^{-}   $ &5955.5$\pm$ 0.4           &       $\frac{3}{2}^{+}$     &5944.05$\pm$ 7.87     &       $\frac{3}{2}^{+}$                     \\
$\Xi _{b}(6227)^{-}   $ &6227.9$\pm$ 0.9$\pm$ 0.2  &       $?^{?}$               &6229.59$\pm$ 1.76     &       $\frac{1}{2}^{-}$           \\
$\Omega _{c}^{0}$       &2695.3$\pm$ 0.4           &       $\frac{1}{2}^{+}$     &2695.19$\pm$ 5.42     &       $\frac{1}{2}^{+}$           \\
$\Omega _{c}(2770)^{0}$ &2766.0                    &       $\frac{3}{2}^{+}$     &2765.59$\pm$ 3.05     &       $\frac{3}{2}^{+}$           \\
$\Omega _{c}(3000)^{0}$ &3000.46$\pm$ 0.25         &       $?^{?}$               &3000.59$\pm$ 1.95     &       $\frac{1}{2}^{-}$           \\
$\Omega _{c}(3050)^{0}$ &3050.17$\pm$ 0.19         &       $?^{?}$               &3050.43$\pm$ 1.76     &       $\frac{1}{2}^{-}$           \\
$\Omega _{c}(3065)^{0}$ &3065.58$\pm$ 0.21         &       $?^{?}$               &3065.79$\pm$ 1.70     &       $\frac{3}{2}^{-}$           \\
$\Omega _{c}(3090)^{0}$ &3090.15$\pm$ 0.26         &       $?^{?}$               &3090.44$\pm$ 1.73     &       $\frac{3}{2}^{-}$           \\
$\Omega _{c}(3120)^{0}$ &3118.98$\pm$ 0.12$\pm$0.23&       $?^{?}$               &3119.32$\pm$ 1.70     &       $\frac{5}{2}^{-}$           \\
$\Omega _{c}(3185)^{0}$ &3185$\pm$ 1.7$\pm$ 0.2    &       $?^{?}$               &3188.98$\pm$ 1.91     &       $\frac{1}{2}^{+}$          \\
$\Omega _{c}(3327)^{0}$ &3327.1$\pm$ 1.2$\pm$ 0.2  &       $?^{?}$               &3323.54$\pm$ 1.84     &       $\frac{1}{2}^{+}$           \\
$\Omega _{b}^{-}      $ &6045.8$\pm$ 0.8           &       $\frac{1}{2}^{+}$     &6045.82$\pm$ 6.53     &       $\frac{1}{2}^{+}$             \\
$\Omega _{b}(6316)^{-}$ &6315.6$\pm$ 0.3$\pm$ 0.5  &       $?^{?}$               &6315.68$\pm$ 2.29     &       $\frac{1}{2}^{-}$          \\
$\Omega _{b}(6330)^{-}$ &6330.3$\pm$ 0.3$\pm$ 0.5  &       $?^{?}$               &6330.08$\pm$ 2.28     &       $\frac{1}{2}^{-}$          \\
$\Omega _{b}(6340)^{-}$ &6339.7$\pm$ 0.3$\pm$ 0.5  &       $?^{?}$               &6341.64$\pm$ 2.28     &       $\frac{3}{2}^{-}$           \\
$\Omega _{b}(6350)^{-}$ &6349.8$\pm$ 0.4$\pm$ 0.5  &       $?^{?}$               &6349.98$\pm$ 2.28     &       $\frac{5}{2}^{-}$           \\
\hline\hline
\end{tabular}
\end{table}

In recent decades, various theoretical approaches are adopted in the investigation of the mass spectrum of the $\Sigma_{Q}$, $\Xi^{\prime}_{Q}$, and $\Omega_{Q}$ baryons, including the quark model \cite{MaltmanIsgur:A11, RobertsPervin:A11, YoshidaOka:A11}, Regge trajectory model \cite{PP:A11, ChenDong:PP888}, relativistic quark model \cite{CapstickIsgur:A11, MiguraMerten:A11, YuWang:A11}, lattice QCD \cite{HBahtiyar:A11, BrownDetmold:A11, JVijande:A11}, relativistic flux tube model \cite{JakhadRai:A11}, chiral quark model \cite{XiaoZhong:A11, YangSegovia:A11}, QCD sum rules \cite{LiuZhuHosaka:A11}, Hypercentral constituent quark model \cite{GandhiRai:A11}, and so on. For more related discussions about these states, we can refer to the Refs. \cite{RRoncaglia:A11, DEbertGalkin:A11, Guerrero:A11, GarciaTecocoatzi:A11, RiveraTecocoatzi:A11, AValcarce:A11}. In Ref. \cite{EbertF:A11}, the authors calculated the mass spectra of ground, orbitally and radially excited heavy baryons in the heavy-quark-light-diquark picture in the framework of the QCD-motivated relativistic quark model. The Ref. \cite{ZhangLuo:A11} further employed the nonrelativistic potential model, combined with the Gaussian expansion method, to calculate the mass spectra and decay behaviors of the experimentally unobserved $1F$-wave singly bottom baryons.

In this study, we investigate the mass spectra of $1F$-wave states for the singly heavy $\Sigma_{Q}$, $\Xi^{\prime}_{Q}$, and $\Omega_{Q}$ baryons, which have yet to be observed in experiments. By analyzing the Regge trajectory formula in the relativistic flux tube model, we get the spin-average masses of the baryons with the orbital and radial quantum number. Based on the relativistic effects, one can obtain the effective masses of the heavy quark and two light quarks. In addition, to obtain the mass shifts, we propose a new scheme of the $6\times6$ matrix elements matrix of state classification and employ the scaling relations to calculate the spin-coupling parameters, and then the mass spectra of $1F$-wave states for the $\Sigma_{Q}$, $\Xi^{\prime}_{Q}$, and $\Omega_{Q}$ baryons are predicted.

This paper is organized as follows. We analyze the Regge trajectory formula to give the spin-average mass of the excited states for the $\Sigma_{Q}$, $\Xi^{\prime}_{Q}$, and $\Omega_{Q}$ baryons in Sec. II. In Sec. III, we study the effective masses of heavy quark and light diquark using the relativistic effect mass formula. In Sec. IV, we review about the spin-dependent Hamiltonian and the scaling relations. The spin-coupling parameters of singly heavy baryons are calculated in Sec. V. In Sec. VI, we discuss the mass spectra of the $\Omega_{c}$ and $\Omega_{b}$ baryons. In Sec. VII, we discuss the mass spectra of the $\Sigma_{c}$ and $\Sigma_{b}$ baryons. In Sec. VIII, we discuss the mass spectra of the $\Xi^{\prime}_{c}$ and $\Xi^{\prime}_{b}$ baryons. Finally, we outline our conclusion in Sec. IX.

\section{The Regge trajectory and spin-average masses}\label{Sec.II}

In the heavy-light diquark picture, the single heavy baryon can be effectively described as a two-body system consisting of a heavy quark and a light diquark. Under this framework, the total mass of the singly heavy baryon state, denoted as $M_{Qqq}$, can be decomposed into a spin-independent term and a spin-dependent mass splitting, expressed as:
\begin{equation}
M_{Qqq}= \bar{M}(n, L) + \Delta M , \label{Mass}
\end{equation}
where $\bar M(n, L)$ is the spin-independent mass depending on the radial quantum number $n$ ($n=0, 1, 2, \cdot\cdot\cdot$) and the orbital quantum number $L$ $(L = 0, 1, 2 ,3, \cdot\cdot\cdot)$, and $\Delta M$ is the mass splitting arising from the spin-dependent interaction $H$.

As a first step toward constructing the mass spectrum of the excited states for the $\Sigma_{Q}$, $\Xi^{\prime}_{Q}$ and $\Omega_{Q}$ baryons, we adopt a linear Regge relation between the mass and angular momentum quantum number in Refs. \cite{PP:A11, LaCourse:A13, ChenWei:A13},
\begin{equation}
(\bar M(L)-M_{Q})^{2}=\pi\alpha L+\left( M_{d}+M_{Q}\left( 1-\frac{m_{\text{cur}Q}^{2}}{M_{Q}^{2}}\right) \right) ^{2},  \label{pp4}
\end{equation}
where $M_{Q}$ and $M_{d}$ are the effective masses of the heavy and light diquark, respectively. $m_{\text{cur}Q}$ can be regarded as current mass of the heavy quark $Q$. Then $\alpha$ stands for the tension of the QCD string connecting the heavy quark at one end and light diquark at the other. In Ref. \cite{EFG:A11}, the slope $\alpha^{\prime}$ of the linear trajectories were extracted for mesons and compared with those previously obtained for light and heavy-light mesons \cite{EFG:AA11, EFG:AAA11}. The slope $\alpha^{\prime}$ was found to decrease rapidly with increasing quark mass, while the slopes for charmonia and $D$, $D_{s}$ mesons, and likewise for bottomonia, $B_{c}$ and $B$, $B_{s}$ mesons, turned out to be very close to each other. This indicates that the slope $\alpha^{\prime}$ of the Regge trajectory for the hadron is mainly determined by the mass $M_{Q}$ of the heavy quark, and in both planes, it generally follows a simple form,
\begin{equation}
\alpha^{\prime}=\frac{1}{2\pi \alpha} \propto \frac{1}{\sqrt{M_{Q}}}.  \label{pt1}
\end{equation}
Thus, the string tension coefficient $\alpha$ is assumed to be proportional to the mass $\sqrt{M_{Q}}$, which is defined as
\begin{eqnarray}
\alpha = \frac{k}{2\pi}(M_{Q})^{\frac{1}{2}},  \label{pptt1}
\end{eqnarray}
where $k$ is a parameter. As a result, the $1/\sqrt{M_{Q}}$ law is essentially a fitting relation based on the Regge trajectories of mesons. When extending this law to the singly heavy baryons considered below, the obtained string tension $\alpha$ should be regarded as an approximate estimate.

Substituting the above Eq. (\ref{pptt1}) into Eq. (\ref{pp4}), we have
\begin{equation}
(\bar M(L)-M_{Q})^{2}=\frac{k}{2}(M_{Q})^{\frac{1}{2}}L+\left( M_{d}+M_{Q}\left( 1-\frac{m_{\text{cur}Q}^{2}}{M_{Q}^{2}}\right) \right) ^{2}.  \label{q666}
\end{equation}
In order to expand to the radial quantum number $n$, we replace $L$ with $L+1.37 n+h$ from our previous work \cite{PP:A11}. Therefore, the Eq. (\ref{q666}) become
\begin{equation}
(\bar M(n, L)-M_{Q})^{2}= \frac{k}{2}(M_{Q})^{\frac{1}{2}} (L+1.37n+h)+\left( M_{d}+M_{Q}\left( 1-\frac{m_{\text{cur}Q}^{2}}{M_{Q}^{2}}\right) \right) ^{2}.  \label{qq66}
\end{equation}
Here, $h$ is a parameter. Eq. (\ref{qq66}) thus establishes Regge-like relation among the mass, $n$ and $L$, with non zero intercept for $n$ = 0 and $L$ = 0 state in ($n$, $L$, $(\bar M(n, L)-M_{Q})^{2}$) plane. This linear dependence of the squared mass $(\bar M(n, L)-M_{Q})^{2}$ on $n$ and $L$ is characteristic of the Regge trajectory phenomenology in hadron spectroscopy, where the intercept reflects the confinement dynamics and the effective quark mass scale.

\section{The effective masses of singly heavy baryons}\label{Sec.III}

Within the framework of quantum chromodynamics, the effective mass of quark varies across different hadronic states, which is due to the dynamic change of the strong interaction. Specifically, this variation in the effective mass of quark depends not only on the orbital angular momentum quantum number, but also on the radial quantum number of the hadron.

In the relativistic flux tube (RFT) model \cite{IwasakiTakagi:AAA11, ChenZh:AAA11}, the whole system of the heavy quark, the light diquark, and flux tube is constantly rotating with angular speed around its center of mass.  The string (flux tube) binds the heavy quark and light diquark together inside the hadron, with the heavy quark and the light diquark located at the two ends of the string, as shown in Fig. 1. Within the RFT model, the Lagrangian of singly heavy baryon can be expressed as
\begin{figure*}[htbp]
\centering
\begin{minipage}[ht]{0.49\textwidth}
\includegraphics[width=6.5cm]{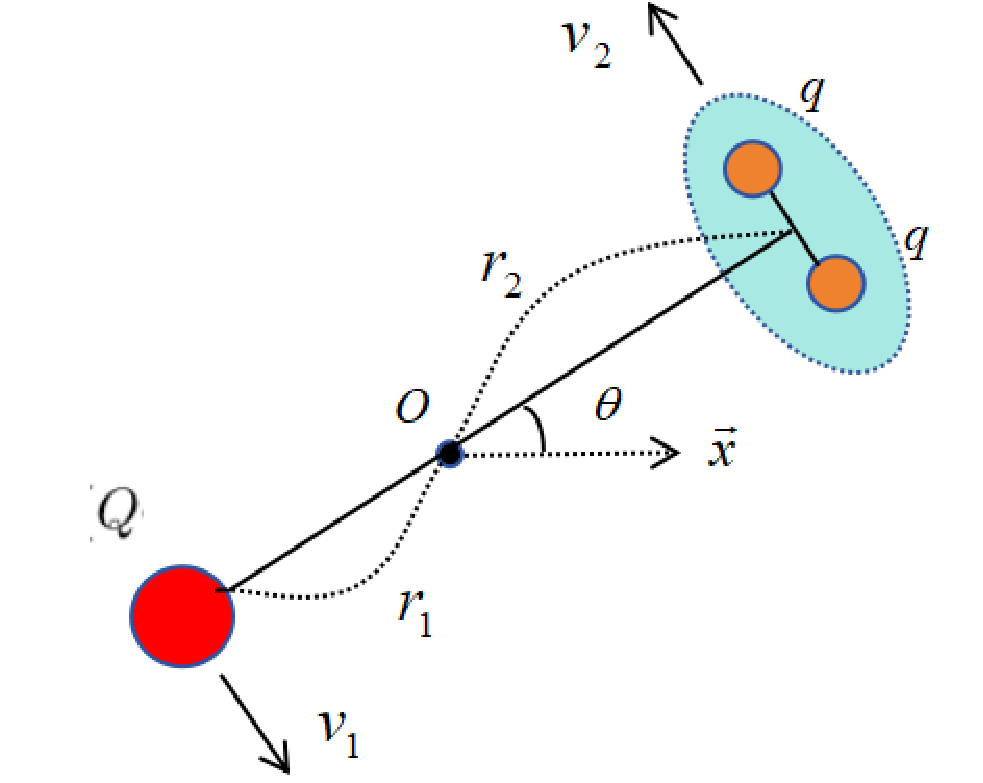}
\caption{Physical picture of the heavy-light quark picture.}
\end{minipage}
\end{figure*}
\begin{equation}
\mathcal{L}(r_{i}, \dot{\theta})= -\sum_{i=1}^{2}\left [m_{i}\sqrt{1-({r_{i}\dot{\theta}})^{2}}+\int_{0}^{r_{i}}\alpha\sqrt{1-({r\dot{\theta}})^{2}}dr\right], \label{ppu2}
\end{equation}
where $m_{i} (i=1,2)$ denote the current masses of the heavy quark and light diquark, respectively. $v_{i}$ = $r_{i}\dot{\theta}$ is the relativistic velocity of the string end heavy quark to which the other light diquark is tied, where $\omega=\dot{\theta}$ stands for the angular velocity of the rotating system. Here, we only consider the transverse velocity of the heavy quark and light antiquark. For the orbital excitations of the singly heavy baryons, the total orbital angular momentum  $L$ = $\partial\mathcal{L}/\partial\dot{\theta}$ is defined by
\begin{equation}
L = \sum_{i=1}^{2}\left[\frac{m_i r_i^{2}\dot{\theta}}{\sqrt{1-(r_i\dot{\theta})^{2}}} + \int_{0}^{r_i} \frac{\alpha r^{2}\dot{\theta}}{\sqrt{1-(r\dot{\theta})^{2}}}\,dr \right] , \label{p8p8p}
\end{equation}
The Hamiltonian $H^{\prime}$ = $\dot{\theta}L$ - $\mathcal{L}$ is then obtained as
\begin{equation}
H^{\prime} = \sum_{i=1}^{2}\left[\frac{m_i}{\sqrt{1-(r_i\dot{\theta})^{2}}} + \int_{0}^{r_i}\frac{\alpha}{\sqrt{1-(r\dot{\theta})^{2}}}\,dr \right].\label{p8p8p8}
\end{equation}
Comparing the above Eqs. (\ref{p8p8p}) and (\ref{p8p8p8}), we can recognize the relativistic structure, in which both the angular momentum and the energy carry a factor of $1/\sqrt{1-v_{i}^{2}}$. It is therefore natural to identify the relativistic masses $M_{1}$ and $M_{2}$ for for the heavy quark and light diquark of the endpoints, respectively,
\begin{eqnarray}
M_{1} &=& \frac{m_{1}}{\sqrt{1-v^{2}_{1}}}, \label{OPP1}      \\
M_{2} &=& \frac{m_{2}}{\sqrt{1-v^{2}_{2}}}, \label{OPP2}
\end{eqnarray}
For simplicity, we have chosen the velocity of light c = 1. Since $M_{1}$ and $M_{2}$ have included the relativistic effect, they might be treated as the effective quark masses.

To estimate the values of the quark velocities $v_{i}$ $(i=1, 2)$ in the heavy-light system we exploit the kinetic energy $T_{i}$ = $\frac{1}{2} m_{i} v_{i}^{2}$ with the current mass $m_{i}$. Then, by taking the average of the square velocity $v_{i}^{2}$, we get
\begin{eqnarray}
\langle v_{i}^{2}\rangle = \langle \frac{2}{m_{i}}T_{i} \rangle. \label{nhf11}
\end{eqnarray}
Here, in the spherical coordinates $r$, $\langle T_{i} \rangle$ can be interpreted as the Virial theorem: $\langle T_{i} \rangle = \frac{1}{2}\langle rV^{\prime}\rangle$. The prime denotes differentiation with respect to $r$. Note that while the Eqs. (\ref{OPP1}) and (\ref{OPP2}) are relativistic of the effective masses, the Eq. (\ref{nhf11}) provides only a non-relativistic estimate for the quark velocities $v_{i}$ $(i=1, 2)$ that appears inside the Eqs. (\ref{OPP1}) and (\ref{OPP2}). This estimate is legitimate, because the velocity for the heavy quark and light diquark in the heavy-light system moves relatively slowly compared with the velocity of light.

Considering the short-range interactions in the three-body quark system, we choose the derivative of the Coulomb potential $V$ = $-4\alpha_{s}/3r$ with the running coupling constant $\alpha_{s}$ of the heavy baryons,
\begin{eqnarray}
\alpha_{s} = \frac{4\pi}{\beta_{0}\ln(\frac{\mu_{Q}^{2}}{\Lambda^{2}})}, \qquad \beta_{0}=(11-\frac{2}{3}n_{f}), \label{OPP101}
\end{eqnarray}
where $n_{f}$ = 3 is a number of flavours and $\Lambda$ is the parameter, and $\mu_{Q}$ = 2$m_{1}m_{2}/(m_{1}+m_{2})$ is the reduced mass of quarks composing the heavy-light diquark system. Combining the momentum conservation $m_{1}v_{1}$ = $m_{2}v_{2}$, the square velocities $\langle v_{i}^{2}\rangle$ with $\langle 1/r\rangle$ = $1/a_{B}N^{2}$ are
\begin{eqnarray}
\langle v_{1}^{2}\rangle &=& \frac{4}{3}\frac{4\pi}{\beta_{0}\ln(\frac{\mu_{Q}^{2}}{\Lambda^{2}})}\frac{1}{m_{1}}\frac{\mu}{N^{2}}, \label{OPP5} \\
\langle v_{2}^{2}\rangle &=& \frac{4}{3}\frac{4\pi}{\beta_{0}\ln(\frac{\mu_{Q}^{2}}{\Lambda^{2}})}\frac{m_{1}}{m^{2}_{2}}\frac{\mu}{N^{2}}, \label{OPP6}
\end{eqnarray}
where $N = n+L+1$ is the principle quantum number of the heavy baryons with $n$ and $L$. $a_{B}$ = $1/\mu$ is the Bohr radius related to the reduced mass $\mu$ = $m_{1}m_{2}/(m_{1}+m_{2})$ in the baryon system. Using the above Eqs. (\ref{OPP5}) and (\ref{OPP6}), Eqs. (\ref{OPP1}) and (\ref{OPP2}) become
\begin{eqnarray}
M_{1} &=& \frac{m_{1}}{\sqrt{1-\frac{4}{3}\frac{4\pi}{\beta_{0}\ln(\frac{\mu_{Q}^{2}}{\Lambda^{2}})}\frac{1}{m_{1}}\frac{\mu}{N^{2}}}}, \label{OPP7} \\
M_{2} &=& \frac{m_{2}}{\sqrt{1-\frac{4}{3}\frac{4\pi}{\beta_{0}\ln(\frac{\mu_{Q}^{2}}{\Lambda^{2}})}\frac{m_{1}}{m^{2}_{2}}\frac{\mu}{N^{2}}}}. \label{OPP8}
\end{eqnarray}

\section{The mass splitting expressions of singly heavy baryons}\label{Sec.IV}

For the orbital excited states of the singly heavy baryons, the mass splitting $\Delta M$ in Eq. (\ref{Mass}) is given by \cite{PP:A11, KarlinerRP:PP888}
\begin{eqnarray}
\Delta M &=& H = a_{1}\mathbf{L}\cdot \mathbf{S}_{d}+a_{2}\mathbf{L}\cdot \mathbf{S}_{Q}+b_{1}S_{12}+c_{1}\mathbf{S}_{d}\cdot \mathbf{S}_{Q},  \label{PP5}
\end{eqnarray}%
where $a_{1}$, $a_{2}$, $b_{1}$, and $c_{1}$ are the spin-coupling parameters. The first two terms are spin-orbit interactions, the third is the tensor energy, and the last is the contact interaction. The spin of heavy quark and light diquark is represented by $\mathbf{S}_{Q}$ and $\mathbf{S}_{d}$, respectively. Further, the tensor energy term is defined as $S_{12}$ = $3(\mathbf{S}_{d}\cdot \mathbf{\hat{r}})(\mathbf{S}_{Q}\cdot \mathbf{\hat{r}})/r^{2}-\mathbf{S}_{d}\cdot \mathbf{S}_{Q}$ with $\hat{r}$ determines the position of the heavy quark relative to the light diquark. Form Refs. \cite{PP:A11} and \cite{KarlinerR:11}, the $S_{12}$ can be expressed as follows
\begin{eqnarray}
S_{12}=-\frac{3}{(2L-1)(2L+3)}[(\mathbf{L}\cdot \mathbf{S}_{d})(\mathbf{L}\cdot \mathbf{S}_{Q})+(\mathbf{L}\cdot \mathbf{S}_{Q})(\mathbf{L}\cdot \mathbf{S}_{d})-\frac{2}{3}L(L+1)(\mathbf{S}_{Q}\cdot \mathbf{S}_{d})].  \label{PP010}
\end{eqnarray}%

In this subsection, we consider the mass splitting $\Delta M$ of the $F$-wave states of the singly heavy baryons with $L=3$. The spin of the diquark $S_{d}=1$ can be coupled with the heavy quark spin $S_{Q}=1/2$ and $L=3$ to the total angular momentum $J$ = $5/2$, $7/2$ or $3/2$, $5/2$, $7/2$, $9/2$. Thus, there are six states in $F$-wave system. Except for the two states of $J=3/2$, $9/2$, the others mix due to the spin-spin interactions. The calculation of the operator $\mathbf{L}\cdot\mathbf{S}_{d}$ and $\mathbf{L}\cdot\mathbf{S}_{Q}$ in Eq. (\ref{PP5}) results in
\begin{equation}
\mathbf{L}\cdot\mathbf{S}_{i} = L_{3}S_{i3}+\left(L_{+}S_{i-}+L_{-}S_{i+}\right)/2,
\end{equation}
with raising and lowering operator $L_{\pm}$, $S_{i\pm}$. In addition, the expectation value of $\mathbf{S}_{d} \cdot \mathbf{S}_{Q}$ in any coupling scheme is
$\langle\mathbf{S}_{d}\cdot \mathbf{S}_{Q}\rangle$ = $[S(S+1)-S_{d}(S_{d}+1)-S_{Q}(S_{Q}+1)]/2$ with the square of the total spin $\mathbf{S}$ = $\mathbf{S}_{d} + \mathbf{S}_{Q}$. In the $L-S$ basis, the expectation values of $\mathbf{L}\cdot\mathbf{S}_{d}$, $\mathbf{L}\cdot\mathbf{S}_{Q}$, $S_{12}$, and $\mathbf{S}_{d}\cdot\mathbf{S}_{Q}$ in Eq. (\ref{PP5}) may be evaluated by explicit construction of states with a given the total angular momentum $J_{3}$ = $S_{d3}+S_{Q3}+L_{3}$, where $S_{d3}$, $S_{Q3}$, and $L_{3}$ represents the third component of $\mathbf{S}_{d}$, $\mathbf{S}_{Q}$, and $\mathbf{L}$, respectively.

For the sake of convenience, we label the $F$-wave states of the baryons by $|^{2S+1} F_{J},J_{3}\rangle$ with linear combinations of the states $|S_{d3}, S_{Q3}, L_{3}\rangle$ of the third components of the respective angular momenta,
\begin{eqnarray}
|^{2S+1} F_{J},J_{3}\rangle = \sum_{S_{d3}, S_{Q3}, L_{3}}\mathcal{C}^{S_{d}, S_{Q}, S}_{S_{d3}, S_{Q3}, S_{3}}\mathcal{C}^{S, L, J}_{S_{3}, L_{3}, J_{3}}|S_{d3}, S_{Q3}, L_{3}\rangle,  \label{P606}
\end{eqnarray}%
where $\mathcal{C}^{S_{d}, S_{Q}, S}_{S_{d3}, S_{Q3}, S_{3}}$, $\mathcal{C}^{S, L, J}_{S_{3}, L_{3}, J_{3}}$ are the Clebsch-Gordan coefficients. With the help of the above relation (\ref{P606}), we can list the $L-S$ coupling basis states for $F$-wave as follows:
\begin{eqnarray}
|^{4} F_{3/2},J_{3}&=&3/2\rangle =\frac{1}{\sqrt{35}}|1,\frac{1}{2},0\rangle-2\sqrt{\frac{2}{105}}|0,\frac{1}{2},1\rangle-\frac{2}{\sqrt{105}}|1,-\frac{1}{2},1\rangle+\sqrt{\frac{2}{21}}|-1,\frac{1}{2},2\rangle  \notag \\
&+&\frac{2}{\sqrt{21}}|0,-\frac{1}{2},2\rangle-\frac{2}{\sqrt{7}}|-1,-\frac{1}{2},3\rangle,
\notag \\
|^{2} F_{5/2},J_{3}&=&5/2\rangle =-\frac{1}{\sqrt{21}}|0,\frac{1}{2},2\rangle+\sqrt{\frac{2}{21}}|1,-\frac{1}{2},2\rangle+\frac{2}{\sqrt{7}}|-1,\frac{1}{2},3\rangle-\sqrt{\frac{2}{7}}|0,-\frac{1}{2},3\rangle,
\notag \\
|^{4} F_{5/2},J_{3}&=&5/2\rangle =\sqrt{\frac{3}{28}}|1,\frac{1}{2},1\rangle-\sqrt{\frac{5}{21}}|0,\frac{1}{2},2\rangle-\sqrt{\frac{5}{42}}|1,-\frac{1}{2},2\rangle+\sqrt{\frac{5}{28}}|-1,\frac{1}{2},3\rangle \notag \\
&+&\sqrt{\frac{5}{14}}|0,-\frac{1}{2},3\rangle,
\notag \\
|^{2} F_{7/2},J_{3}&=&7/2\rangle =-\sqrt{\frac{1}{3}}|0,\frac{1}{2},3\rangle+\sqrt{\frac{2}{3}}|1,-\frac{1}{2},3\rangle,
\notag \\
|^{4} F_{7/2},J_{3}&=&7/2\rangle =\frac{1}{\sqrt{3}}|1,\frac{1}{2},2\rangle-\frac{2}{3}|0,\frac{1}{2},3\rangle-\frac{\sqrt{2}}{3}|1,-\frac{1}{2},3\rangle,
\notag \\
|^{4} F_{9/2},J_{3}&=&9/2\rangle =|1,\frac{1}{2},3\rangle. \label{VV}
\end{eqnarray}%
Given that the six basis states are derived from the specific configuration $|S_{d3}, S_{Q3}, L_{3}\rangle$, we introduce a new scheme of state classification, namely the $6\times6$ matrix elements matrix of the operator $H_{6\times6}$. Therefore, by employing this model, we calculate the expectation values of  $\mathbf{L}\cdot\mathbf{S}_{d}$, $\mathbf{L}\cdot\mathbf{S}_{Q}$, $S_{12}$, and $\mathbf{S}_{d}\cdot\mathbf{S}_{Q}$ as follows,
\begin{eqnarray}
\setlength{\abovedisplayskip}{4pt}
\setlength{\belowdisplayskip}{4pt}
\langle\mathbf{L\cdot S}_{d}\rangle_{6\times6} &=&\left[
\begin{array}{cccccc}
        -4          &          0              &                    0         &                  0    &       0    &    0  \\
         0          &    -\frac{8}{3}         &     -\frac{2\sqrt{5}}{3}     &                  0    &       0    &    0  \\
         0          &  -\frac{2\sqrt{5}}{3}   &         -\frac{7}{3}         &                  0    &       0    &    0  \\
         0          &          0              &                    0         &                  2    & -\sqrt{3}  &    0  \\
         0          &          0              &                    0         &           -\sqrt{3}   &       0    &    0  \\
         0          &          0              &                    0         &                   0   &       0    &    3
\end{array}
\right], \\
\langle\mathbf{L\cdot S}_{Q}\rangle_{6\times6} &=&\left[
\begin{array}{cccccc}
        -2          &          0              &                    0         &                  0    &       0    &    0  \\
         0          &     \frac{2}{3}         &      \frac{2\sqrt{5}}{3}     &                  0    &       0    &    0  \\
         0          &   \frac{2\sqrt{5}}{3}   &         -\frac{7}{6}         &                  0    &       0    &    0  \\
         0          &          0              &                    0         &       -\frac{1}{2}    & \sqrt{3}   &    0  \\
         0          &          0              &                    0         &           \sqrt{3}    &       0    &    0  \\
         0          &          0              &                    0         &                   0   &       0    &   \frac{3}{2}
\end{array}\right], \\
\langle\mathbf{S_{12}}\rangle_{6\times6} &=&\left[
\begin{array}{cccccc}
    -\frac{4}{5}    &          0              &                    0         &                  0       &       0                &    0  \\
         0          &          0              &      \frac{1}{\sqrt{5}}      &                  0       &       0                &    0  \\
         0          &   \frac{1}{\sqrt{5}}    &        \frac{1}{5}           &                  0       &       0                &    0  \\
         0          &          0              &                    0         &                  0       & -\frac{1}{2\sqrt{3}}   &    0  \\
         0          &          0              &                    0         &  -\frac{1}{2\sqrt{3}}    &     \frac{2}{3}        &    0  \\
         0          &          0              &                    0         &                   0      &       0                &   -\frac{1}{3}
\end{array}
\right],\\
\langle\mathbf{S_{d}\cdot S}_{Q}\rangle_{6\times6} &=&\left[
\begin{array}{cccccc}
   \frac{1}{2}      &          0              &                    0         &                  0    &      0        &    0  \\
         0          &         -1              &                    0         &                  0    &      0        &    0  \\
         0          &          0              &           \frac{1}{2}        &                  0    &      0        &    0  \\
         0          &          0              &                    0         &                 -1    &      0        &    0  \\
         0          &          0              &                    0         &                   0   &  \frac{1}{2}  &    0 \\
         0          &          0              &                    0         &                   0   &      0        &   \frac{1}{2}
\end{array}
\right].
\end{eqnarray}
The matrix forms of these mass shifts in Eq. (\ref{PP5}) are
\begin{eqnarray}
\Delta M = H_{6\times6} =
\left[
\begin{array}{cccccc}
   H_{11}   &     0        &     0     &     0    &      0     &   0   \\
      0     &    H_{22}    &   H_{23}  &     0    &      0     &   0  \\
      0     &    H_{32}    &   H_{33}  &     0    &      0     &   0   \\
      0     &     0        &     0     &  H_{44}  &   H_{45}   &   0   \\
      0     &     0        &     0     &  H_{54}  &   H_{55}   &   0  \\
      0     &     0        &     0     &     0    &      0     &   H_{66}
\end{array}
\right],  \label{MMaac12}
\end{eqnarray}
where
\begin{eqnarray}
H_{11} &=& -4a_{1}-2a_{2}-\frac{4}{5}b_{1}+\frac{1}{2}c_{1}, \notag \\
H_{22} &=& -\frac{8}{3}a_{1}+\frac{2}{3}a_{2}-c_{1},  \notag \\
H_{23} &=& H_{32} = -\frac{2\sqrt{5}}{3}a_{1}+\frac{2\sqrt{5}}{3}a_{2}+\frac{\sqrt{5}}{5}b_{1},  \notag \\
H_{33} &=& -\frac{7}{3}a_{1}-\frac{7}{6}a_{2}+\frac{1}{5}b_{1}+\frac{1}{2}c_{1},  \notag \\
H_{44} &=& 2a_{1}-\frac{1}{2}a_{2}-c_{1},  \notag \\
H_{45} &=& H_{54} = -\sqrt{3}a_{1}+\sqrt{3}a_{2}-\frac{\sqrt{3}}{6}b_{1},  \notag \\
H_{55} &=& \frac{2}{3}b_{1}+\frac{1}{2}c_{1}, \notag \\
H_{66} &=& 3a_{1}+\frac{3}{2}a_{2}-\frac{1}{3}b_{1}+\frac{1}{2}c_{1}. \notag
\end{eqnarray}
Diagonalizing the above matrices Eq. (\ref{MMaac12}), one can compute the mass shifts $\Delta M(J,j)$ with the total angular momentum $\mathbf{J}$ and the total light diquark angular momentum $\mathbf{j}=\mathbf{L}+\mathbf{S}_{d}$, where ${S}_{d}=1$ is the spin of the diquark, so $j=2, 3, 4$,
\begin{eqnarray}
\Delta M(3/2,2)&=&-4a_{1}-2a_{2}-\frac{4}{5}b_{1}+\frac{1}{2}c_{1}, \notag \\
\Delta M(5/2,2)&=&\frac{1}{20}\left(-50a_{1}-5a_{2}+2b_{1}-5c_{1}\right) \notag \\
               &-&\frac{1}{60}\sqrt{(90a_{1}-95a_{2}-26b_{1}+5c_{1})^{2}+5(20a_{2}-4b_{1}-20c_{1})^{2}},  \notag \\
\Delta M(5/2,3)&=&\frac{1}{20}\left(-50a_{1}-5a_{2}+2b_{1}-5c_{1}\right) \notag \\
               &+&\frac{1}{60}\sqrt{(90a_{1}-95a_{2}-26b_{1}+5c_{1})^{2}+5(20a_{2}-4b_{1}-20c_{1})^{2}},  \notag \\
\Delta M(7/2,3)&=&\frac{1}{12}\left(12a_{1}-3a_{2}+4b_{1}-3c_{1}\right) \notag \\
               &-&\frac{1}{24}\sqrt{(48a_{1}-39a_{2}+2b_{1}-9c_{1})^{2}+3(9a_{2}-6b_{1}-9c_{1})^{2}},  \notag \\
\Delta M(7/2,4)&=&\frac{1}{12}\left(12a_{1}-3a_{2}+4b_{1}-3c_{1}\right) \notag \\
               &+&\frac{1}{24}\sqrt{(48a_{1}-39a_{2}+2b_{1}-9c_{1})^{2}+3(9a_{2}-6b_{1}-9c_{1})^{2}},  \notag \\
\Delta M(9/2,4)&=&3a_{1}+\frac{3}{2}a_{2}-\frac{1}{3}b_{1}+\frac{1}{2}c_{1}.  \label{MM121}
\end{eqnarray}%

This expresses six mass shifts in terms of four unknown variables parameters. If  all four variables parameters, in general, can be investigated, one can determined the six mass shift states in $F$-wave for the singly heavy baryon system.

\section{The spin-coupling parameters of singly heavy baryons}\label{Sec.V}

In this section, let us first examine the mass scaling of the spin couplings between singly heavy charmed and bottom baryons. To elaborate on the mass shifts $\Delta M(J, j)$ in Eq. (\ref{MM121}) for the entire baryon system, we employ the following scaling relations to describe the spin-coupling parameters $a_{1}$, $a_{2}$, $b_{1}$, and $c_{1}$ given by \cite{PP:A11, Pan:A11},
\begin{equation}
\left\{
\begin{array}{rrrr} \vspace{1ex}
a_{1}(B_{a}, (n+1)L)&=&\frac{M^{\prime}_{Q}M^{\prime}_{d}}{M_{Q}M_{d}}\frac{{N^{\prime}_{a_{1}}}}{{N_{a_{1}}}}a_{1}(B_{a}^{\prime}, (n^{\prime}+1)L^{\prime}), \\ \vspace{1ex}
a_{2}(B_{a}, (n+1)L)&=&\frac{M^{\prime}_{Q}M^{\prime}_{d}}{M_{Q}M_{d}}\frac{{N^{\prime}_{a_{2}}}}{{N_{a_{2}}}}a_{2}(B_{a}^{\prime}, (n^{\prime}+1)L^{\prime}), \\  \vspace{1ex}
b_{1}(B_{a}, (n+1)L)&=&\frac{M^{\prime}_{Q}M^{\prime}_{d}}{M_{Q}M_{d}}\frac{{N^{\prime}_{b_{1}}}}{{N_{b_{1}}}}b_{1}(B_{a}^{\prime}, (n^{\prime}+1)L^{\prime}), \\
c_{1}(B_{a}, (n+1)L)&=&\frac{M^{\prime}_{Q}M^{\prime}_{d}}{M_{Q}M_{d}}\frac{{N^{\prime}_{c_{1}}}}{{N_{c_{1}}}}c_{1}(B_{a}^{\prime}, (n^{\prime}+1)L^{\prime}),\\
\end{array}%
\right.   \label{scr:PP888}
\end{equation}%
where $M^{\prime}_{Q}$, $M_{Q}$ are the effective masses of heavy quarks, and $M^{\prime}_{d}$, $M_{d}$ are the effective masses of light diquark in the singly heavy baryon system. $n, n^{\prime}=0, 1 , 2, \cdots$, $L, L^{\prime}= S, P, D, F, \cdots$. $B_{a}, B_{a}^{\prime}$ represent baryons with
\begin{eqnarray}
N_{a_{1}} &=& (n+L+1)^{2} = N_{a_{2}}, \notag \\
N_{b_{1}} &=& L(L+1/2)(L+1)(n+L+1)^{3}, \notag \\
N_{c_{1}} &=& (L+\lambda)(n+L+1)^{3}, \label{mm11}
\end{eqnarray}
corresponding to the similar form of $N^{\prime}_{a_{1}}$, $N^{\prime}_{a_{2}}$, $N^{\prime}_{b_{1}}$, and $N^{\prime}_{c_{1}}$ with $L^{\prime}$ and $n^{\prime}$, respectively, where the prime denotes the quantities of the baryon $B_{a}^{\prime}$ obtained from experiments, distinguishing them from that of an unobserved baryon $B_{a}$.

In our previous work \cite{PP:A11}, based on the experimental data of both the LHCb \cite{Aaij:A11} and Belle collaborations \cite{YeltonBel:PP888}, we have calculated the mass spectra of all singly heavy baryons using the Regge trajectory model. Obviously, the five excited states of $\Omega_{c}(3000)^{0}$, $\Omega_{c}(3050)^{0}$, $\Omega_{c}(3065)^{0}$, $\Omega_{c}(3090)^{0}$, and $\Omega_{c}(3120)^{0}$) for the $\Omega_{c}$ baryons are determined as the $1P$ wave states corresponding to $J^{P}=1/2^{-}$, $1/2^{-}$, $3/2^{-}$, $3/2^{-}$, and $5/2^{-}$, respectively. In addition, in Ref. \cite{Ali:PP888}, the authors fixed the values of the spin-coupling parameters $a_{1}$, $a_{2}$, $b_{1}$, and $c_{1}$ for the $\Omega_{c}$ baryons in Table \ref{Table:PP861} with the uncertainties using the method of least squares. The same result can be found in Ref. \cite{KarlinerR:11} for narrow excited $\Omega_{c}$ baryons.  By applying Eq. (\ref{scr:PP888}), we can obtain the parameters $a_{1}$, $a_{2}$, $b_{1}$, and $c_{1}$ for the partner $\Sigma_{c}$ and $\Xi^{\prime}_{c}$ baryons as shown in Table \ref{Table:PP861}, respectively.
\renewcommand{\tabcolsep}{0.7cm}
\renewcommand{\arraystretch}{1.0}
\begin{table}[tbh]
\caption{ The parameters in $1P$-wave states (in MeV) of singly heavy $\Sigma_{c}$, $\Xi_{c}^{\prime}$, and $\Omega_{c}$ baryons. \label{Table:PP861}}%
\label{tab:Eff-mass}
\begin{tabular}
[c]{ccccc}\hline\hline
\text{baryon} & $a_{1}$ & $a_{2}$ & $b_{1}$ & $c_{1}$ \\\hline
$\Omega_{c}$ & 26.96 $\pm$ 0.28 & 25.76 $\pm$ 0.76 & 13.51 $\pm$ 0.54 & 4.04 $\pm$ 0.44   \\\hline
$\Sigma_{c}$ & 35.86 $\pm$ 0.37 & 34.27 $\pm$ 1.01 & 17.97 $\pm$ 0.72 & 5.37 $\pm$ 0.59    \\\hline
$\Xi^{\prime}_{c}$  & 30.64 $\pm$ 0.32 & 29.28 $\pm$ 0.86 & 15.35 $\pm$ 0.62 & 4.59 $\pm$ 0.50     \\\hline\hline
\end{tabular}
\end{table}

To elaborate on the mass shifts $\Delta M(J, j)$ in Eq. (\ref{MM121}) for the entire baryon system, these parameters in Table \ref{Table:PP861} for the $1P$-wave $\Omega_{c}$ states are employed as the object of the scaling relations (\ref{scr:PP888}) to calculate the parameters of the other states. The current masses of the quarks are extracted from PDG \cite{Navas:PP888} as follows
\begin{eqnarray}
&m_{u} = 0.00216\ \text{GeV}, \  m_{s} = 0.0935\ \text{GeV}, \notag  \\
&m_{c} = 1.273\ \text{GeV}, \ m_{b} = 4.183\ \text{GeV}.  \label{VVV000}
\end{eqnarray}
By utilizing the Eqs. (\ref{OPP8}) and (\ref{VVV000}), the effective masses of the $1F$-wave states are obtained as
\begin{eqnarray}
&M_{uu} = 0.0044\ \text{GeV}, \  M_{su} = 0.0987\ \text{GeV},  \notag   \\
&M_{ss} = 0.193\ \text{GeV}, \  M_{c} = 1.274\ \text{GeV}, \ M_{b} = 4.184\ \text{GeV}.  \label{VVV333}
\end{eqnarray}
Due to the influence of the confinement effect, the effective masses of quarks may differ significantly from their current masses. These results indicate that the effective mass of a quark is not a fixed constant. Instead, it varies with the orbital quantum number $L$, the radial quantum number $n$ of the quark inside the hadron. Consequently, the masses $M_{1}$ and $M_{2}$ in Eqs. (\ref{OPP7}) and (\ref{OPP8}) decrease monotonically with increasing the square of $N$ with $L$ and $n$. In other words, the effective masses are largest in the ground state ($S$-wave) and decrease gradually toward the current masses as the orbital excitation proceeds from $P$-wave to $D$-wave and to $F$-wave.

There are two fitting parameters, $h$ and $\lambda$, appearing in Eqs. (\ref{qq66}) and (\ref{mm11}), respectively. By analyzing the experimental values, we take the values $h$ = 1.72 and $\lambda$ = 0.9. In the ground state system, all the others in Eq. (\ref{PP5}) vanish except for the last term, which becomes dominant and therefore governs the low energy behavior. For the higher excited states, the parameters $b_{1}$ and $c_{1}$ are relatively small and can thus be disregarded in the present approximation. In Eq. (\ref{pptt1}), the parameter $k$ is determined from an analysis of the experimental data provided by PDG \cite{Navas:PP888}, we have
\begin{eqnarray}
&&k(\Omega_{c}) = 2.094\ \text{GeV}, \; k(\Sigma_{c}) = 1.540\ \text{GeV}, \; k(\Xi^{\prime}_{c}) = 1.770\ \text{GeV},  \notag   \\
&&k(\Omega_{b}) = 1.652\ \text{GeV}, \; k(\Sigma_{b}) = 0.973\ \text{GeV}, \;  k(\Xi^{\prime}_{b}) = 1.541\ \text{GeV}. \label{jj21}
\end{eqnarray}
Meanwhile, the running coupling constant $\alpha_{s}$ is considered with $\Lambda$ in Eq. (\ref{OPP101}),
\begin{eqnarray}
&\Lambda(\Omega_{c}) = 0.083\ \text{GeV}, \alpha_{s}(\Omega_{c}) = 0.510\ \text{GeV}, \ \Lambda(\Omega_{b}) = 0.138\ \text{GeV}, \alpha_{s}(\Omega_{b}) = 0.734\ \text{GeV},  \notag   \\
&\Lambda(\Sigma_{c}) = 0.003\ \text{GeV}, \alpha_{s}(\Sigma_{c}) = 0.618\ \text{GeV}, \ \Lambda(\Sigma_{b}) = 0.003\ \text{GeV}, \alpha_{s}(\Sigma_{b}) = 0.661\ \text{GeV},  \notag   \\
&\Lambda(\Xi^{\prime}_{c}) = 0.065\ \text{GeV}, \alpha_{s}(\Xi^{\prime}_{c}) = 0.691\ \text{GeV}, \ \Lambda(\Xi^{\prime}_{b}) = 0.070\ \text{GeV}, \alpha_{s}(\Xi^{\prime}_{b}) = 0.749\ \text{GeV}.  \label{jj12}
\end{eqnarray}
With the help of the above parameters (\ref{VVV000})-(\ref{jj12}), we calculate the spin-independent mass $\bar M$ (\ref{qq66}) and the parameters $a_{1}$, $a_{2}$, $b_{1}$, and $c_{1}$ of the $1F$-wave states for the singly heavy $\Sigma_{Q}$, $\Xi_{Q}^{\prime}$, and $\Omega_{Q}$ baryons. The results are presented in Table \ref{Table:PP8866}. While the uncertainties in the parameters of the $\Sigma_{Q}$, $\Xi_{Q}^{\prime}$, and $\Omega_{Q}$ baryons, which are determined from the error results in Table \ref{Table:PP8866} by using the scaling relations (\ref{scr:PP888}).
\renewcommand{\tabcolsep}{0.7cm}
\renewcommand{\arraystretch}{1.0}
\begin{table}[tbh]
\caption{ The parameters in $1F$-wave states (in MeV) of singly heavy $\Sigma_{Q}$, $\Xi_{Q}^{\prime}$, and $\Omega_{Q}$ baryons. \label{Table:PP8866}}%
\label{tab:Eff-mass}
\begin{tabular}
[c]{cccccc}\hline\hline
\text{baryon}&  $a_{1}$ & $a_{2}$ & $b_{1}$ & $c_{1}$ \\\hline
$\Omega_{c}$ &  7.18 $\pm$ 0.07 & 6.86 $\pm$ 0.20 & 0.13 $\pm$ 0.005 & 0.26 $\pm$ 0.03   \\
$\Omega_{b}$ &  2.16 $\pm$ 0.03 & 2.06 $\pm$ 0.09 & 0.04 $\pm$ 0.002 & 0.08 $\pm$ 0.009   \\
$\Sigma_{c}$ &  9.79 $\pm$ 0.10 & 9.36 $\pm$ 0.28 & 0.18 $\pm$ 0.007 & 0.36 $\pm$ 0.04    \\
$\Sigma_{b}$ &  2.98 $\pm$ 0.03 & 2.84 $\pm$ 0.08 & 0.06 $\pm$ 0.002& 0.11 $\pm$ 0.01    \\
$\Xi^{\prime}_{c}$ &  8.41 $\pm$ 0.09 & 8.04 $\pm$ 0.24 & 0.15 $\pm$ 0.006 & 0.31 $\pm$ 0.03     \\
$\Xi^{\prime}_{b}$ &  2.55 $\pm$ 0.03 & 2.44 $\pm$ 0.08 & 0.05 $\pm$ 0.002 & 0.09 $\pm$ 0.01     \\\hline\hline
\end{tabular}
\end{table}

\section{The $\Omega_{c}$ and $\Omega_{b}$ baryons}\label{Sec.VI}

For $\Omega_{c}$ baryon, which contains a charmed quark $c$ and two light strange quarks $ss$, in which the two light quarks are often treated as a diquark. From Table \ref{Table 1}, we found that the calculated masses for $S$-, $P$- and $D$-wave excited states are closer to the experimentally measured masses. Although these baryons are well established in experiment, their quantum numbers are still unknown apart from the ground states.

In 2017, LHCb Collaboration discovered five new narrow $\Omega_{c}$ states observed in decay channel $\Xi^{+}_{c}K$ \cite{Aaij:A11}, defined as $\Omega_{c}(3000)^{0}$, $\Omega_{c}(3050)^{0}$, $\Omega_{c}(3065)^{0}$, $\Omega_{c}(3090)^{0}$, and $\Omega_{c}(3120)^{0}$. In 2021, the LHCb collaboration analyzed the helical angle distributions of $\Omega_{c}(3000)^{0}$, $\Omega_{c}(3050)^{0}$, $\Omega_{c}(3065)^{0}$, $\Omega_{c}(3090)^{0}$ in $\Omega^{-}_{b}$ $\rightarrow$ $\Xi_{c}^{+}K^{-}\pi^{-}$ decays \cite{Aaij:PPPP888}. Their analysis disfavor the $J$ = $1/2$ hypothesis for $\Omega_{c}(3050)$ and $\Omega_{c}(3065)$ at the $2.2\sigma$ and $3.6\sigma$ level, respectively, and rejects rejects the combined assignment of $J^{P}=1/2^{-}$, $1/2^{-}$, $3/2^{-}$ and $3/2^{-}$ at $3.5\sigma$ standard deviations, while the parity remains undetermined. In Ref. \cite{ZhangN:PP888}, the authors perform a systematic analysis of the low-lying $\Omega_{c}$ structures in a coupled-channel approach. The absence of $\Omega_{c}(3000)^{0}$  and $\Omega_{c}(3050)^{0}$ can be classified as the lower states with $J^{P}=1/2^{-}$ and $3/2^{-}$, respectively. They assign $J^{P}=3/2^{-}$ and $5/2^{-}$ for the $\Omega_{c}(3065)^{0}$ and $\Omega_{c}(3090)^{0}$ states, respectively. While the $\Omega_{c}(3120)^{0}$ may correspond to a higher pole with quantum number $J^{P}=3/2^{-}$. In our model, we propose that all of these five observed states \cite{Aaij:A11} of $\Omega_{c}$ baryon belong to the $1P$ wave states with $J^{P}$ = $1/2^{-}$, $1/2^{-}$, $3/2^{-}$, $3/2^{-}$, and $5/2^{-}$, respectively. Additionally, $\Omega_{c}(3185)^{0}$ and $\Omega_{c}(3327)^{0}$, could be grouped into the $2S$- and $1D$-wave charmed baryon family with $J^{P}$ = $1/2^{+}$ and $1/2^{+}$, respectively. The mixed state $\Omega_c(3327)^0$ has been speculated as a $2S$ state in Ref. \cite{KaerRoter:PP888} and as a $1D$ state in Refs. \cite{LuoLiu:PP888}. However, we still need more observable objects to get clarity about the internal structure. In addition, in Ref. \cite{CL:A11} the authors suggested that $\Omega_{c}(3185)^{0}$ may be regarded as a $2S$ state with $J^{P}=1/2^{+}$ or $J^{P}=3/2^{+}$, or their overlapping structure, and $\Omega_{c}(3185)^{0}$ is interpreted as a $P$-wave state in Ref. \cite{PachecoBijker:PP888}. This prediction will definitely be helpful for future experiments to detect these unobserved states. For more information of $\Omega_{c}$ baryons, we can refer to Refs. \cite{HYCheng:PP888, JOudichhya:PP888, SuenagaOka:PP888, PatelShah:PP888, LiuLuo:PP888}.

Within our theoretical framework, we calculate the spin-independent mass of the $1F$-wave states for the $\Omega_{c}$ baryon with $n$ = 0, $L$ = 3 by using Eq. (\ref{qq66}),
\begin{eqnarray}
\bar M(0, 3)(\Omega_{c}) &=& M_{c}+\sqrt{\frac{k}{2}(M_{c})^{\frac{1}{2}} (L+1.37n+h)+\left( M_{ss}+M_{c}\left( 1-\frac{m_{\text{cur}c}^{2}}{M_{c}^{2}}\right) \right) ^{2}} \notag  \\
             &=& 3642.81 \text{MeV}. \label{qqmm5566}
\end{eqnarray}
The parameters are
\begin{eqnarray}
a_{1}(\Omega_{c},1F)&=&\frac{M^{\prime}_{c}M^{\prime}_{ss}}{M_{c}M_{ss}}\frac{(n^{\prime}+L^{\prime}+1)^{2}}{(n+L+1)^{2}}a_{1}(\Omega_{c},1P)=7.18\ \text{MeV}, \label{wemm681}\\
a_{2}(\Omega_{c},1F)&=&\frac{M^{\prime}_{c}M^{\prime}_{ss}}{M_{c}M_{ss}}\frac{(n^{\prime}+L^{\prime}+1)^{2}}{(n+L+1)^{2}}a_{2}(\Omega_{c},1P)=6.86\ \text{MeV}, \label{wemm682}\\
b_{1}(\Omega_{c},1F)&=&\frac{M^{\prime}_{c}M^{\prime}_{ss}}{M_{c}M_{ss}}\frac{L^{\prime}(L^{\prime}+\frac{1}{2})(L^{\prime}+1)(n^{\prime}+L^{\prime}+1)^{3}}{L(L+\frac{1}{2})(L+1)(n+L+1)^{3}}b_{1}(\Omega_{c},1P)=0.13\ \text{MeV}, \label{wemm683}\\
c_{1}(\Omega_{c},1F)&=&\frac{M^{\prime}_{c}M^{\prime}_{ss}}{M_{c}M_{ss}}\frac{(L^{\prime}+0.9)(n^{\prime}+L^{\prime}+1)^{3}}{(L+0.9)(n+L+1)^{3}}c_{1}(\Omega_{c},1P)=0.26\ \text{MeV}. \label{wemm684}
\end{eqnarray}
By substituting the spin-average mass (\ref{qqmm5566}) and the parameters in Eqs. (\ref{wemm681})-(\ref{wemm684}) into Eqs. (\ref{MM121}) and (\ref{Mass}), we can obtain the values of masses of the $1F$-wave excited states for the $\Omega_{c}$ baryon. The mass results are listed in Table \ref{ppdd55}. In our approach, the calculated mass spectra of $\Omega_{c}(1^{4}F, 3/2^{-})$, $\Omega_{c}(1^{2}F, 5/2^{-})$, $\Omega_{c}(1^{4}F, 5/2^{-})$, $\Omega_{c}(1^{2}F, 7/2^{-})$, $\Omega_{c}(1^{4}F, 7/2^{-})$, and $\Omega_{c}(1^{4}F, 9/2^{-})$ fall within the range of approximately 3600.42 MeV to 3674.70 MeV. Thus, the mass splitting between $\Omega_{c}({\small 1}^{2}{\small F}_{3/2})$ and $\Omega_{c}({\small 1}^{4}{\small F}_{9/2})$ states is $\Delta M$ = $M(\Omega_{c}({\small 1}^{4}{\small F}_{9/2}))$ $-$ $M(\Omega_{c}({\small 1}^{2}{\small F}_{3/2}))$ = 74.28 MeV. In relativistic quark-diquark picture \cite{EbertF:A11}, its mass splitting is much larger than the result 26.28 MeV of our model calculation. While the mass of the lowest $1F$-wave state is $M(\Omega_{c}, {\small 1}^{4}{\small F}_{3/2})$ = 3600.42 MeV, which is larger than 273.32 MeV that of $D$-wave $\Omega_{c}(3327)^{0}$ state. As a comparison, the predictions from other models are also listed in Table \ref{ppdd55}.

For $\Omega_{b}$ baryon, composed of one bottom quark $b$ and two strange quarks $ss$, is the heaviest state in the singly heavy baryon family. In 2020, LHCb collaboration has reported four narrow states in the $\Xi_{b}^{0} K^{-}$ spectrum \cite{Aaij:A12}, defined as $\Omega _{b}(6316)^{-}$, $\Omega _{b}(6330)^{-}$, $\Omega _{b}(6340)^{-}$, and $\Omega _{b}(6350)^{-}$. According to our discussion with the observations, the four states may be assigned as the predicted five $1P$-wave excitations with $J^{P}=1/2^{-}$, $1/2^{-}$, $3/2^{-}$ and $3/2^{-}$, respectively. A similar result can be referred to in Ref. \cite{KarlinerR:12}. Ref. \cite{XuHuang:12} consider all $P$-wave states represented by interpolating currents with a derivative and calculate the corresponding masses residues using the QCD sum rule method. They suggested the $\Omega _{b}(6316)^{-}$ is likely a $P$-wave $\Omega_{c}$ baryon with either $J^{P}=1/2^{-}$ or $3/2^{-}$, associated with a $\rho$-mode excitation. The $\Omega _{b}(6330)^{-}$, $\Omega _{b}(6340)^{-}$ are identified as partner states with $\lambda$-mode excitations, having $J^{P}=1/2^{-}$ and $1/2^{-}$, respectively. The $\Omega _{b}(6350)^{-}$ state is to be a $P$-wave $\Omega _{b}$ baryon with $J^{P}=1/2^{-}$ and a $\lambda$-mode excitation. We have interpreted them as the the predicted five $1P$-wave excitations with $J^{P}$ = $1/2^{-}$, $1/2^{-}$, $3/2^{-}$, and $5/2^{-}$, see Table \ref{Table 1}. For the $1F$-wave states of the $\Omega_{b}$ baryon, according to this information in our model, the spin-average mass is given by using Eq. (\ref{qq66}) with $n$ = 0, $L$ = 3,
\begin{eqnarray}
\bar M(0, 3)(\Omega_{b}) &=& M_{b}+\sqrt{\frac{k}{2}(M_{b})^{\frac{1}{2}} (L+1.37n+h)+\left( M_{ss}+M_{b}\left( 1-\frac{m_{\text{cur}b}^{2}}{M_{b}^{2}}\right) \right) ^{2}} \notag  \\
             &=& 7013.85 \text{MeV}. \label{qqmm6666}
\end{eqnarray}
The parameters are
\begin{eqnarray}
a_{1}(\Omega_{b},1F)&=&\frac{M^{\prime}_{c}M^{\prime}_{ss}}{M_{b}M_{ss}}\frac{(n^{\prime}+L^{\prime}+1)^{2}}{(n+L+1)^{2}}a_{1}(\Omega_{c},1P)=2.16\ \text{MeV}, \label{wemm661}\\
a_{2}(\Omega_{b},1F)&=&\frac{M^{\prime}_{c}M^{\prime}_{ss}}{M_{b}M_{ss}}\frac{(n^{\prime}+L^{\prime}+1)^{2}}{(n+L+1)^{2}}a_{2}(\Omega_{c},1P)=2.06\ \text{MeV}, \label{wemm662}\\
b_{1}(\Omega_{b},1F)&=&\frac{M^{\prime}_{c}M^{\prime}_{ss}}{M_{b}M_{ss}}\frac{L^{\prime}(L^{\prime}+\frac{1}{2})(L^{\prime}+1)(n^{\prime}+L^{\prime}+1)^{3}}{L(L+\frac{1}{2})(L+1)(n+L+1)^{3}}b_{1}(\Omega_{c},1P)=0.04\ \text{MeV}, \label{wemm663}\\
c_{1}(\Omega_{b},1F)&=&\frac{M^{\prime}_{c}M^{\prime}_{ss}}{M_{b}M_{ss}}\frac{(L^{\prime}+0.9)(n^{\prime}+L^{\prime}+1)^{3}}{(L+0.9)(n+L+1)^{3}}c_{1}(\Omega_{c},1P)=0.08\ \text{MeV}. \label{wemm664}
\end{eqnarray}
In the same way, by substituting the spin-average mass (\ref{qqmm6666}) and the parameters in Eqs. (\ref{wemm661})-(\ref{wemm664}) into Eqs. (\ref{MM121}) and (\ref{Mass}), we can obtain the values of masses of the $\Omega_{b}$ baryon. The results are also listed in Table \ref{ppdd55} and compared with different models. According to our result in Table \ref{ppdd55}, the predicted masses of the six states $\Omega_{b}(1^{4}F, 3/2^{-})$, $\Omega_{b}(1^{2}F, 5/2^{-})$, $\Omega_{b}(1^{4}F, 5/2^{-})$, $\Omega_{b}(1^{2}F, 7/2^{-})$, $\Omega_{b}(1^{4}F, 7/2^{-})$, and $\Omega_{b}(1^{4}F, 9/2^{-})$ lie between 7001.10 MeV and 7023.45 MeV. Consequently, the mass splitting $M(\Omega_{b}(1^{4}F, 9/2^{-}))$ $-$ $M(\Omega_{b}(1^{4}F, 3/2^{-}))$ = 22.35 MeV between $\Omega_{b}(1^{4}F, 3/2^{-})$ and $\Omega_{b}(1^{4}F, 9/2^{-})$ states is much smaller than the result 51.93 for the $\Omega_{c}$ baryon states. The mass differences between these six states for $\Omega_{b}$ baryon become relatively small due to the existence of heavy bottom quark. For the $J^{P}$ = $5/2^{-}$ case, the two $\Omega_{b}(1^{2}F, 5/2^{-})$ and $\Omega_{b}(1^{4}F, 5/2^{-})$ states are the mass degenerate mixed states in $1F$-wave baryons, or mixed with the first state $\Omega_{b}(1^{4}F, 3/2^{-})$. They are most likely consistent with a single resonance because its degeneracy. However, it is actually composed of these two states. Similarly, the two $\Omega_{b}(1^{2}F, 7/2^{-})$ and $\Omega_{b}(1^{4}F, 7/2^{-})$ states with $J^{P}$ = $7/2^{-}$ are also close to each other or mixed with the third state of $J^{P}$ = $5/2^{-}$.

In addition, to compute the uncertainties of the mass spectra of the $\Omega_{c}$ and $\Omega_{b}$ baryons in \ref{ppdd55}, we considered all possible sources of the uncertainties. Based on the error analysis of the experimental value in Table \ref{Table 1}, as well as considering the errors of the spin-coupling parameters $a_{1}$, $a_{2}$, $b_{1}$, and $c_{1}$ in \ref{Table:PP8866} and the errors of other input parameters. Therefore, the uncertainties in the mass spectra of the $\Omega_{c}$ and $\Omega_{b}$ baryons are estimated by adding these error results in quadrature.

\renewcommand{\tabcolsep}{0.55cm}
\renewcommand{\arraystretch}{1.0}
\begin{table}[htbp]
\caption{Mass spectra (MeV) of $\Omega_{Q}$ baryons are given and compared with different quark models.}\label{ppdd55}
{\begin{tabular}{cccccc}
\hline\hline
{\small baryon } {\small State } $J^{P}$  &{Ours}&   \cite{EbertF:A11}   & \cite{PengLiu:PP888}  &  \cite{YuWang:A11}  &  \cite{JakhadRai:A11, ShahRai:A11} \\
\hline
$%
\begin{array}{rrr}
\Omega_{c}&{\small 1}^{4}{\small F}_{3/2} & {\small 3/2}^{-} \\
\Omega_{c}&{\small 1}^{2}{\small F}_{5/2} & {\small 5/2}^{-} \\
\Omega_{c}&{\small 1}^{4}{\small F}_{5/2} & {\small 5/2}^{-} \\
\Omega_{c}&{\small 1}^{2}{\small F}_{7/2} & {\small 7/2}^{-} \\
\Omega_{c}&{\small 1}^{4}{\small F}_{7/2} & {\small 7/2}^{-} \\
\Omega_{c}&{\small 1}^{4}{\small F}_{9/2} & {\small 9/2}^{-}%
\end{array}%
$ & $%
\begin{array}{r}
{\small 3600.42 \pm 1.92} \\
{\small 3618.20 \pm 1.87} \\
{\small 3628.00 \pm 1.86}\\
{\small 3642.99 \pm 1.85} \\
{\small 3653.50 \pm 1.86}\\
{\small 3674.70 \pm 1.88}%
\end{array}%
$ & $%
\begin{array}{r}
{\small 3533} \\
{\small 3522} \\
{\small 3515} \\
{\small 3514} \\
{\small 3498} \\
{\small 3485}%
\end{array}%
$ & $%
\begin{array}{r}
{\small 3540} \\
{\small 3546} \\
{\small 3532} \\
{\small 3536} \\
{\small 3521} \\
{\small 3521}%
\end{array}%
$ & $%
\begin{array}{r}
{\small 3525} \\
{\small 3528} \\
{\small 3525} \\
{\small 3529} \\
{\small 3524} \\
{\small 3529}%
\end{array}%
$ & $%
\begin{array}{r}
{\small 3457} \\
{\small 3403} \\
{\small 3418}\\
{\small 3369} \\
{\small 3354}\\
{\small 3310}%
\end{array}%
$ \\
\hline

$
\begin{array}{rrr}
\Omega_{b}&{\small 1}^{4}{\small F}_{3/2} & {\small 3/2}^{-} \\
\Omega_{b}&{\small 1}^{2}{\small F}_{5/2} & {\small 5/2}^{-} \\
\Omega_{b}&{\small 1}^{4}{\small F}_{5/2} & {\small 5/2}^{-} \\
\Omega_{b}&{\small 1}^{2}{\small F}_{7/2} & {\small 7/2}^{-} \\
\Omega_{b}&{\small 1}^{4}{\small F}_{7/2} & {\small 7/2}^{-} \\
\Omega_{b}&{\small 1}^{4}{\small F}_{9/2} & {\small 9/2}^{-}%
\end{array}%
$ & $%
\begin{array}{r}
{\small 7001.10 \pm 2.72} \\
{\small 7006.45 \pm 2.71} \\
{\small 7009.40 \pm 2.71}\\
{\small 7013.91 \pm 2.71} \\
{\small 7017.07 \pm 2.71}\\
{\small 7023.45 \pm 2.72}%
\end{array}%
$ & $%
\begin{array}{r}
{\small 6763} \\
{\small 6771} \\
{\small 6737} \\
{\small 6736} \\
{\small 6719} \\
{\small 6713}%
\end{array}%
$ & $%
\begin{array}{r}
{\small 6753} \\
{\small 6758} \\
{\small 6742} \\
{\small 6747} \\
{\small 6728} \\
{\small 6731}%
\end{array}%
$ & $%
\begin{array}{r}
{\small 6762} \\
{\small 6767} \\
{\small 6751} \\
{\small 6755} \\
{\small 6736} \\
{\small 6739}%
\end{array}%
$ & $%
\begin{array}{r}
{\small 6877.0} \\
{\small 6885.5} \\
{\small 6921.1} \\
{\small 6979.8} \\
{\small 6929.0} \\
{\small 6986.5}%
\end{array}%
$ \\
\hline\hline
\end{tabular}}
\end{table}

\section{The $\Sigma_{c}$ and $\Sigma_{b}$ baryons}\label{Sec.VII}

For $\Sigma_{c}$ baryon, the $1S$-wave states with $J^{P}$ = $1/2^{+}$ and $3/2^{+}$ have been very well established experimentally. Belle Collaboration observed that $\Sigma_{c}(2800)$ state decay into $\Lambda_{c}^{+}\pi$, which might be a good candidate of $1P$-wave excitations. In this work, we assign $J^{P}$ = $3/2^{-}$ to $\Sigma_{c}(2800)$. Our calculation results are also very close to the experimental measurements, as summarized in Table \ref{Table 1}. Despite this progress, the properties of higher excited states, such as $1D$- and $1F$-wave states, remain unclear due to the lack of sufficient experimental data. For the $1F$-wave states of $\Sigma_{c}$ baryon, one can use Eq. (\ref{qq66}) with $n$ = 0 and $L$ = 3 to obtain the spin-independent mass $\bar M$,
\begin{eqnarray}
\bar M(0, 3)(\Sigma_{c}) &=& M_{c}+\sqrt{\frac{k}{2}(M_{c})^{\frac{1}{2}} (L+1.37n+h)+\left( M_{uu}+M_{c}\left( 1-\frac{m_{\text{cur}c}^{2}}{M_{c}^{2}}\right) \right) ^{2}} \notag  \\
             &=& 3298.00 \text{MeV}, \label{qq660}
\end{eqnarray}
The parameters are
\begin{eqnarray}
a_{1}(\Sigma_{c},1F)&=&\frac{M^{\prime}_{c}M^{\prime}_{uu}}{M_{c}M_{uu}}\frac{(n^{\prime}+L^{\prime}+1)^{2}}{(n+L+1)^{2}}a_{1}(\Sigma_{c},1P)=9.79\ \text{MeV},\\
a_{2}(\Sigma_{c},1F)&=&\frac{M^{\prime}_{c}M^{\prime}_{uu}}{M_{c}M_{uu}}\frac{(n^{\prime}+L^{\prime}+1)^{2}}{(n+L+1)^{2}}a_{2}(\Sigma_{c},1P)=9.36\ \text{MeV},\\
b_{1}(\Sigma_{c},1F)&=&\frac{M^{\prime}_{c}M^{\prime}_{uu}}{M_{c}M_{uu}}\frac{L^{\prime}(L^{\prime}+\frac{1}{2})(L^{\prime}+1)(n^{\prime}+L^{\prime}+1)^{3}}{L(L+\frac{1}{2})(L+1)(n+L+1)^{3}}b_{1}(\Sigma_{c},1P)=0.18\ \text{MeV},\\
c_{1}(\Sigma_{c},1F)&=&\frac{M^{\prime}_{c}M^{\prime}_{uu}}{M_{c}M_{uu}}\frac{(L^{\prime}+0.9)(n^{\prime}+L^{\prime}+1)^{3}}{(L+0.9)(n+L+1)^{3}}c_{1}(\Sigma_{c},1P)=0.36\ \text{MeV}. \label{we1132}
\end{eqnarray}
According to Table \ref{ppddmd55}, the values of mass spectra for all six $\Sigma_{c}$ states increase with the increase of the spin parity quantum numbers $J^{P}$ = $3/2^{-}$, $5/2^{-}$, $5/2^{-}$, $7/2^{-}$, $7/2^{-}$, and $9/2^{-}$. The calculated masses of $\Sigma_{c}({\small 1}^{4}{\small F}_{3/2})$ and $\Sigma_{c}({\small 1}^{4}{\small F}_{9/2})$ states with $J^{P}=3/2^{-}$ and $9/2^{-}$ are projected to be 3240.14 MeV and 3341.54 MeV, respectively. The mass splitting value between them is 101.40 MeV, which is the largest among the baryons containing light diquark $uu$. Consequently, the total mass splitting between these six states is approximately 20 MeV. In comparison, in relativistic quark-diquark picture \cite{EbertF:A11}, the mass spectrum shows a decreasing trend with increasing $J^{P}$, which is contrary to our model calculation. This difference demonstrates the importance of the theoretical framework in describing the mass spectrum of singly heavy baryons.

In the case of $\Sigma_{b}$ baryon, $\Sigma_{b}(6097)$ has been measured using fully reconstructed $\Lambda^{0}_{b}\rightarrow\Lambda^{+}_{c}\pi^{-}$ and $\Lambda^{+}_{c}\rightarrow p K^{-}_{c}\pi^{+}$ decays in Ref. \cite{collaborationT:PPP888}. In our calculations, $\Sigma_{b}(6097)$, can be a good candidate of $1P$-wave excitations with $J^{P}=3/2^{-}$. However, the $1F$-wave $\Sigma_{b}$ states have not been observed yet. In this study, we provide predictions for the masses of the $1F$-wave $\Sigma_{b}$ states, which could be instrumental in the search for higher orbital excited $\Sigma_{b}$ states by experiment. By employing Eqs. (\ref{qq66}) and (\ref{scr:PP888}) with $n$ = 0 and $L$ = 3, we also calculate the spin-independent mass of the $\Sigma_{b}$ baryons,
\begin{eqnarray}
\bar M(0, 3)(\Sigma_{b}) &=& M_{b}+\sqrt{\frac{k}{2}(M_{b})^{\frac{1}{2}} (L+1.37n+h)+\left( M_{uu}+M_{b}\left( 1-\frac{m_{\text{cur}b}^{2}}{M_{b}^{2}}\right) \right) ^{2}} \notag  \\
             &=& 6566.59 \text{MeV}, \label{qq6630}
\end{eqnarray}
The parameters are
\begin{eqnarray}
a_{1}(\Sigma_{b},1F)&=&\frac{M^{\prime}_{c}M^{\prime}_{uu}}{M_{b}M_{uu}}\frac{(n^{\prime}+L^{\prime}+1)^{2}}{(n+L+1)^{2}}a_{1}(\Sigma_{c},1P)=2.98\ \text{MeV},\\
a_{2}(\Sigma_{b},1F)&=&\frac{M^{\prime}_{c}M^{\prime}_{uu}}{M_{b}M_{uu}}\frac{(n^{\prime}+L^{\prime}+1)^{2}}{(n+L+1)^{2}}a_{2}(\Sigma_{c},1P)=2.84\ \text{MeV},\\
b_{1}(\Sigma_{b},1F)&=&\frac{M^{\prime}_{c}M^{\prime}_{uu}}{M_{b}M_{uu}}\frac{L^{\prime}(L^{\prime}+\frac{1}{2})(L^{\prime}+1)(n^{\prime}+L^{\prime}+1)^{3}}{L(L+\frac{1}{2})(L+1)(n+L+1)^{3}}b_{1}(\Sigma_{c},1P)=0.06\ \text{MeV},\\
c_{1}(\Sigma_{b},1F)&=&\frac{M^{\prime}_{c}M^{\prime}_{uu}}{M_{b}M_{uu}}\frac{(L^{\prime}+0.9)(n^{\prime}+L^{\prime}+1)^{3}}{(L+0.9)(n+L+1)^{3}}c_{1}(\Sigma_{c},1P)=0.11\ \text{MeV}. \label{we1132}
\end{eqnarray}
In our calculation model, we present our mass predictions of the $1F$-wave baryon states in Table \ref{ppddmd55} for $\Sigma_{b}$ baryon, and the spin-independent mass of 6566.59 MeV for the  $\Sigma_{b}$ baryon is higher than 3268.59 MeV of the $\Sigma_{c}$ baryon given in Eq. (\ref{qq660}). The mass splitting between all six states is found to be approximately 10-15 MeV, which is smaller than 10 MeV for the $\Sigma_{c}$ baryon due to the existence of the large mass of heavy quark $b$. As shown in Table \ref{ppddmd55}, the $\Sigma_{b}({\small 1}^{2}{\small F}_{5/2})$ and $\Sigma_{b}({\small 1}^{4}{\small F}_{5/2})$ states are the mass degenerate mixed states in $1F$-wave with $J^{P}$ = $5/2^{-}$. Similarly, the $\Sigma_{b}({\small 1}^{2}{\small F}_{7/2})$ and $\Sigma_{b}({\small 1}^{4}{\small F}_{7/2})$ states are also the mass degenerate mixed states with $J^{P}$ = $7/2^{-}$. These predicted masses are slightly higher than those reported in Ref. \cite{EbertF:A11}. Our results may serve as a crucial reference for experimentalists in identifying higher orbital excitations in the $\Sigma_{c}$ and $\Sigma_{b}$ baryons. Additionally, the principal sources of uncertainties in the masses of the $\Sigma_{c}$ and $\Sigma_{b}$ states are due to the uncertainties with respect to the auxiliary parameters and errors of other input parameters.

\renewcommand{\tabcolsep}{0.55cm}
\renewcommand{\arraystretch}{1.0}
\begin{table}[htbp]
\caption{Mass spectra (MeV) of $\Sigma_{Q}$ baryons are given and compared with different quark models.}\label{ppddmd55}
{\begin{tabular}{ccccccc}
\hline\hline
{\small baryon }\; {\small State }\; $J^{P}$  &{Ours}&   \cite{EbertF:A11}   & \cite{PengLiu:PP888}  &  \cite{YuWang:A11}  &  \cite{JakhadRai:A11, ShahRai:A11} \\
\hline
$%
\begin{array}{rrr}
\Sigma_{c}&{\small 1}^{4}{\small F}_{3/2} & {\small 3/2}^{-} \\
\Sigma_{c}&{\small 1}^{2}{\small F}_{5/2} & {\small 5/2}^{-} \\
\Sigma_{c}&{\small 1}^{4}{\small F}_{5/2} & {\small 5/2}^{-} \\
\Sigma_{c}&{\small 1}^{2}{\small F}_{7/2} & {\small 7/2}^{-} \\
\Sigma_{c}&{\small 1}^{4}{\small F}_{7/2} & {\small 7/2}^{-} \\
\Sigma_{c}&{\small 1}^{4}{\small F}_{9/2} & {\small 9/2}^{-}%
\end{array}%
$ & $%
\begin{array}{r}
{\small 3240.14 \pm 2.20} \\
{\small 3264.41 \pm 2.11} \\
{\small 3277.79 \pm 2.10}\\
{\small 3298.25 \pm 2.08} \\
{\small 3312.59 \pm 2.10}\\
{\small 3341.54 \pm 2.16}%
\end{array}%
$ & $%
\begin{array}{r}
{\small 3288} \\
{\small 3283} \\
{\small 3254} \\
{\small 3253} \\
{\small 3227} \\
{\small 3209}%
\end{array}%
$ & $%
\begin{array}{r}
{\small 3277} \\
{\small 3283} \\
{\small 3247} \\
{\small 3252} \\
{\small 3208} \\
{\small 3209}%
\end{array}%
$ & $%
\begin{array}{r}
{\small 3299} \\
{\small 3304} \\
{\small 3299} \\
{\small 3305} \\
{\small 3299} \\
{\small 3305}%
\end{array}%
$ & $%
\begin{array}{r}
{\small 3332} \\
{\small 3245} \\
{\small 3268}\\
{\small 3189} \\
{\small 3165}\\
{\small 3094}%
\end{array}%
$ \\
\hline

$
\begin{array}{rrr}
\Sigma_{b}&{\small 1}^{4}{\small F}_{3/2} & {\small 3/2}^{-} \\
\Sigma_{b}&{\small 1}^{2}{\small F}_{5/2} & {\small 5/2}^{-} \\
\Sigma_{b}&{\small 1}^{4}{\small F}_{5/2} & {\small 5/2}^{-} \\
\Sigma_{b}&{\small 1}^{2}{\small F}_{7/2} & {\small 7/2}^{-} \\
\Sigma_{b}&{\small 1}^{4}{\small F}_{7/2} & {\small 7/2}^{-} \\
\Sigma_{b}&{\small 1}^{4}{\small F}_{9/2} & {\small 9/2}^{-}%
\end{array}%
$ & $%
\begin{array}{r}
{\small 6549.02 \pm 2.78} \\
{\small 6556.39 \pm 2.77} \\
{\small 6560.45 \pm 2.77}\\
{\small 6566.67 \pm 2.77} \\
{\small 6571.03 \pm 2.77}\\
{\small 6579.82 \pm 2.78}%
\end{array}%
$ & $%
\begin{array}{r}
{\small 6550} \\
{\small 6564} \\
{\small 6501} \\
{\small 6500} \\
{\small 6472} \\
{\small 6459}%
\end{array}%
$ & $%
\begin{array}{r}
{\small 6539} \\
{\small 6544} \\
{\small 6508} \\
{\small 6512} \\
{\small 6466} \\
{\small 6469}%
\end{array}%
$ & $%
\begin{array}{r}
{\small 6538} \\
{\small 6542} \\
{\small 6506} \\
{\small 6510} \\
{\small 6463} \\
{\small 6465}%
\end{array}%
$ & $%
\begin{array}{r}
{\small 6558.7} \\
{\small 6567.2} \\
{\small 6624.6} \\
{\small 6712.3} \\
{\small 6632.2} \\
{\small 6718.6}%
\end{array}%
$ \\
\hline\hline
\end{tabular}}
\end{table}

\section{The $\Xi^{\prime}_{c}$ and $\Xi^{\prime}_{b}$ baryons}\label{Sec.VIII}

For the orbital excited $\Xi^{\prime}_{c}$ and $\Xi^{\prime}_{b}$ baryons, the established candidates include the $1S$-, $1P$-, and $1D$-wave states. The measured masses are listed Table \ref{Table 1}. In contrast, the $1F$-wave excitations for both $\Xi^{\prime}_{c}$ and $\Xi^{\prime}_{b}$ baryons have not yet been observed in any experimental facility to date. In this section, we present the systematic predictions for the masses of unobserved $1F$-wave $\Xi^{\prime}_{c}$ and $\Xi^{\prime}_{b}$ states, which could be instrumental in the search for higher orbital excited states by experiment. In Tables \ref{Table 1}, it can be seen that the model predictions for these $S$-wave $\Xi^{\prime}_{c}$ and $\Xi^{\prime}_{b}$ states with $J^{P}$ = $1/2^{+}$, $3/2^{+}$ also agree well with the experimental data, respectively. The previously observed $\Xi_{c}(2923)$, $\Xi_{c}(2930)$, and $\Xi_{b}(6227)$ states are good candidates for the $P$-wave states. We assign $J^{P}$ = $3/2^{-}$, $3/2^{-}$ and $1/2^{-}$ to these three states, respectively. For the $D$-wave state, we assign $J^{P}$ = $1/2^{+}$ to $\Xi_{c}(3123)$ state. Our calculated masses for these states are very close to the experimental values.

A similar method can be applied to the excited $\Xi^{\prime}_{c}$ and $\Xi^{\prime}_{b}$ baryon systems in order to analyze their masses and parameters. Specifically, we also calculate the spin-independent mass of the $\Xi^{\prime}_{c}$ and $\Xi^{\prime}_{b}$ baryons with $n$ = 0, $L$ = 3, respectively,
\begin{eqnarray}
\bar M(0, 3)(\Xi^{\prime}_{c}) &=& M_{c}+\sqrt{\frac{k}{2}(M_{c})^{\frac{1}{2}} (L+1.37n+h)+\left( M_{su}+M_{c}\left( 1-\frac{m_{\text{cur}c}^{2}}{M_{c}^{2}}\right) \right) ^{2}} \notag  \\
             &=& 3446.47 \text{MeV}. \label{qq6630}
\end{eqnarray}
and
\begin{eqnarray}
\bar M(0, 3)(\Xi^{\prime}_{b}) &=& M_{b}+\sqrt{\frac{k}{2}(M_{b})^{\frac{1}{2}} (L+1.37n+h)+\left( M_{su}+M_{b}\left( 1-\frac{m_{\text{cur}b}^{2}}{M_{b}^{2}}\right) \right) ^{2}} \notag  \\
             &=& 6912.15 \text{MeV}. \label{qqmm66}
\end{eqnarray}
The parameters are
\begin{eqnarray}
a_{1}(\Xi^{\prime}_{c},1F)&=&\frac{M^{\prime}_{c}M^{\prime}_{su}}{M_{c}M_{su}}\frac{(n^{\prime}+L^{\prime}+1)^{2}}{(n+L+1)^{2}}a_{1}(\Xi^{\prime}_{c},1P)=8.41\ \text{MeV},\\
a_{2}(\Xi^{\prime}_{c},1F)&=&\frac{M^{\prime}_{c}M^{\prime}_{su}}{M_{c}M_{su}}\frac{(n^{\prime}+L^{\prime}+1)^{2}}{(n+L+1)^{2}}a_{2}(\Xi^{\prime}_{c},1P)=8.04\ \text{MeV},\\
b_{1}(\Xi^{\prime}_{c},1F)&=&\frac{M^{\prime}_{c}M^{\prime}_{su}}{M_{c}M_{su}}\frac{L^{\prime}(L^{\prime}+\frac{1}{2})(L^{\prime}+1)(n^{\prime}+L^{\prime}+1)^{3}}{L(L+\frac{1}{2})(L+1)(n+L+1)^{3}}b_{1}(\Xi^{\prime}_{c},1P)=0.15\ \text{MeV},\\
c_{1}(\Xi^{\prime}_{c},1F)&=&\frac{M^{\prime}_{c}M^{\prime}_{su}}{M_{c}M_{su}}\frac{(L^{\prime}+0.9)(n^{\prime}+L^{\prime}+1)^{3}}{(L+0.9)(n+L+1)^{3}}c_{1}(\Xi^{\prime}_{c},1P)=0.31\ \text{MeV}. \label{we1132}
\end{eqnarray}
and
\begin{eqnarray}
a_{1}(\Xi^{\prime}_{b},1F)&=&\frac{M^{\prime}_{c}M^{\prime}_{su}}{M_{b}M_{su}}\frac{(n^{\prime}+L^{\prime}+1)^{2}}{(n+L+1)^{2}}a_{1}(\Xi^{\prime}_{c},1P)=2.55\ \text{MeV},\\
a_{2}(\Xi^{\prime}_{b},1F)&=&\frac{M^{\prime}_{c}M^{\prime}_{su}}{M_{b}M_{su}}\frac{(n^{\prime}+L^{\prime}+1)^{2}}{(n+L+1)^{2}}a_{2}(\Xi^{\prime}_{c},1P)=2.44\ \text{MeV},\\
b_{1}(\Xi^{\prime}_{b},1F)&=&\frac{M^{\prime}_{c}M^{\prime}_{su}}{M_{b}M_{su}}\frac{L^{\prime}(L^{\prime}+\frac{1}{2})(L^{\prime}+1)(n^{\prime}+L^{\prime}+1)^{3}}{L(L+\frac{1}{2})(L+1)(n+L+1)^{3}}b_{1}(\Xi^{\prime}_{c},1P)=0.05\ \text{MeV},\\
c_{1}(\Xi^{\prime}_{b},1F)&=&\frac{M^{\prime}_{c}M^{\prime}_{su}}{M_{b}M_{su}}\frac{(L^{\prime}+0.9)(n^{\prime}+L^{\prime}+1)^{3}}{(L+0.9)(n+L+1)^{3}}c_{1}(\Xi^{\prime}_{c},1P)=0.09\ \text{MeV}. \label{we1132}
\end{eqnarray}
Hence, the predicted masses for the $\Xi^{\prime}_{c}$ and $\Xi^{\prime}_{b}$ states are calculated. The results are listed in Table \ref{ppddmm55}. The lowest $1F$-wave state with $J^{P}$ = $3/2^{-}$ is predicted at 3396.78 MeV. This is approximately 120 MeV above the assigned $1D$ state $\Xi_{c}(3123)$, making it a promising candidate for experimental search due to its proximity to the $1D$ states. For $\Xi^{\prime}_{c}$ states, with $J^{P}$ = $3/2^{-}$, $5/2^{-}$, $5/2^{-}$, $7/2^{-}$, $7/2^{-}$, and $9/2^{-}$, the mass splittings range from 11.48  MeV to 24.86 MeV. While they are significantly higher than 3.48 MeV to 7.53 MeV for $\Xi^{\prime}_{b}$ states, reflecting the suppression of spin-dependent effects in the heavier bottom baryon due to the $1/M_{Q}$ scaling of hyperfine interactions in the heavy quark limit. Similar to $\Omega_{Q}$ and $\Sigma_{Q}$ baryons, we employ the auxiliary parameters as well as the errors of other input parameters to compute the uncertainties in the masses of the $\Xi^{\prime}_{Q}$ states. Therefore, our results may serve as a crucial reference for experimentalists in identifying higher orbital excitations in the $\Xi^{\prime}_{c}$ and $\Xi^{\prime}_{b}$ baryons.
\renewcommand{\tabcolsep}{0.55cm}
\renewcommand{\arraystretch}{1.0}
\begin{table}[htbp]
\caption{Mass spectra (MeV) of $\Xi^{\prime}_{Q}$ baryons are given and compared with different quark models.}\label{ppddmm55}
{\begin{tabular}{ccccccc}
\hline\hline
{\small baryon }\;  {\small State }\;  $J^{P}$  &{Ours}&    \cite{EbertF:A11}   & \cite{PengLiu:PP888}  &  \cite{WangLi:PP888}  &  \cite{JakhadRai:A11, ShahRai:A11} \\
\hline
$%
\begin{array}{rrr}
\Xi^{\prime}_{c}&{\small 1}^{4}{\small F}_{3/2} & {\small 3/2}^{-} \\
\Xi^{\prime}_{c}&{\small 1}^{2}{\small F}_{5/2} & {\small 5/2}^{-} \\
\Xi^{\prime}_{c}&{\small 1}^{4}{\small F}_{5/2} & {\small 5/2}^{-} \\
\Xi^{\prime}_{c}&{\small 1}^{2}{\small F}_{7/2} & {\small 7/2}^{-} \\
\Xi^{\prime}_{c}&{\small 1}^{4}{\small F}_{7/2} & {\small 7/2}^{-} \\
\Xi^{\prime}_{c}&{\small 1}^{4}{\small F}_{9/2} & {\small 9/2}^{-}%
\end{array}%
$ & $%
\begin{array}{r}
{\small 3396.78 \pm 2.06} \\
{\small 3417.63 \pm 1.99} \\
{\small 3429.11 \pm 1.99}\\
{\small 3446.68 \pm 1.96} \\
{\small 3459.00 \pm 1.98}\\
{\small 3483.86 \pm 2.03}%
\end{array}%
$ & $%
\begin{array}{r}
{\small 3418} \\
{\small 3408} \\
{\small 3394} \\
{\small 3393} \\
{\small 3373} \\
{\small 3357}%
\end{array}%
$ & $%
\begin{array}{r}
{\small 3427} \\
{\small 3433} \\
{\small 3408} \\
{\small 3412} \\
{\small 3383} \\
{\small 3383}%
\end{array}%
$ & $%
\begin{array}{r}
{\small 3424} \\
{\small 3428} \\
{\small 3424} \\
{\small 3428} \\
{\small 3423} \\
{\small 3428}%
\end{array}%
$ & $%
\begin{array}{r}
{\small 3332} \\
{\small 3245} \\
{\small 3268} \\
{\small 3189} \\
{\small 3165} \\
{\small 3094}%
\end{array}%
$ \\
\hline

$
\begin{array}{rrr}
\Xi^{\prime}_{b}&{\small 1}^{4}{\small F}_{3/2} & {\small 3/2}^{-} \\
\Xi^{\prime}_{b}&{\small 1}^{2}{\small F}_{5/2} & {\small 5/2}^{-} \\
\Xi^{\prime}_{b}&{\small 1}^{4}{\small F}_{5/2} & {\small 5/2}^{-} \\
\Xi^{\prime}_{b}&{\small 1}^{2}{\small F}_{7/2} & {\small 7/2}^{-} \\
\Xi^{\prime}_{b}&{\small 1}^{4}{\small F}_{7/2} & {\small 7/2}^{-} \\
\Xi^{\prime}_{b}&{\small 1}^{4}{\small F}_{9/2} & {\small 9/2}^{-}%
\end{array}%
$ & $%
\begin{array}{r}
{\small 6897.08 \pm 2.08} \\
{\small 6903.40 \pm 2.08} \\
{\small 6906.88 \pm 2.07}\\
{\small 6912.21 \pm 2.07} \\
{\small 6915.95 \pm 2.07}\\
{\small 6923.48 \pm 2.08}%
\end{array}%
$ & $%
\begin{array}{r}
{\small 6675} \\
{\small 6686} \\
{\small 6640} \\
{\small 6641} \\
{\small 6619} \\
{\small 6610}%
\end{array}%
$ & $%
\begin{array}{r}
{\small 6649} \\
{\small 6654} \\
{\small 6629} \\
{\small 6633} \\
{\small 6601} \\
{\small 6604}%
\end{array}%
$ & $%
\begin{array}{r}
{\small 6657} \\
{\small 6660} \\
{\small 6657} \\
{\small 6660} \\
{\small 6657} \\
{\small 6661}%
\end{array}%
$ & $%
\begin{array}{r}
{\small 6728.8} \\
{\small 6737.3} \\
{\small 6781.4} \\
{\small 6851.4} \\
{\small 6789.2} \\
{\small 6858.0}%
\end{array}%
$ \\
\hline\hline
\end{tabular}}
\end{table}

\section{Summary and outlook}\label{Sec.IX}

In this work, we focus on the spectroscopic analysis of the experimentally unobserved $1F$-wave states for the singly heavy $\Sigma_{Q}$, $\Xi^{\prime}_{Q}$, and $\Omega_{Q}$ baryons. The mass of the baryons consists of the sum of two parts: spin-independent mass $\bar M(n, L)$ and mass splitting $\Delta M$. For the spin-independent masses of the baryons, in order to obtain the spin-independent mass $\bar M(n, L)$, we use a linear Regge relation between the mass and angular momentum quantum number to complete the spin-average masses of the baryons. Meanwhile, we employ the relativistic effective mass formula under the Coulomb potential to study the effective masses of the heavy quark and light diquark.

For the spin-dependent masses $\Delta M$, we consider the Hamiltonian $H$ = $a_{1}\mathbf{L}\cdot \mathbf{S}_{d}+a_{2}\mathbf{L}\cdot \mathbf{S}_{Q}+b_{1}S_{12}+c_{1}\mathbf{S}_{d}\cdot \mathbf{S}_{Q}$ with the operator $\mathbf{L}\cdot\mathbf{S}_{d}$, $\mathbf{L}\cdot\mathbf{S}_{Q}$, $S_{12}$, and $\mathbf{S}_{d} \cdot \mathbf{S}_{Q}$. Diagonalizing the $6\times6$ matrix forms of the Hamiltonian $H$, one can obtain the forms of six mass shifts in terms of four unknown variables parameters $a_{1}$, $a_{2}$, $b_{1}$, and $c_{1}$. Based on chromodynamics similarity between the baryons, we adopt the scaling relationship to calculate these spin-coupling parameters. Our calculation results indicate that the $1F$-wave states of singly heavy baryons exhibit interesting mass spectrometric characteristics.

With our calculation model, we investigate the mass spectra of $1^{4}F_{3/2}$, $1^{2}F_{5/2}$, $1^{4}F_{5/2}$, $1^{2}F_{7/2}$, $1^{4}F_{7/2}$, and $1^{4}F_{9/2}$ states for the $\Sigma_{Q}$, $\Xi^{\prime}_{Q}$, and $\Omega_{Q}$ baryons. It can be seen from Table \ref{Table 1} that our predictions for the mass of $S$-, $P$-, and $D$-wave states are very close to the experimental measurement values. Although the masses of the excited $\Sigma_{Q}$, $\Xi^{\prime}_{Q}$, and $\Omega_{Q}$ baryon states have not yet been experimentally discovered, our estimations of the $1F$-wave states in Table \ref{ppddmm55}, \ref{ppddmd55}, and \ref{ppdd55} are more reliable. Finally, we hope that these results may provide important references for the experiment, such as LHCb, Belle, $BABAR$, and CLEO, and contribute to a deeper understanding of the spectroscopy of singly heavy $\Sigma_{Q}$, $\Xi^{\prime}_{Q}$, and $\Omega_{Q}$ baryons.

\end{document}